\documentclass[12pt,a4paper]{article}
\usepackage{jheppub}
\usepackage{amsmath,amssymb,amsfonts,mathtools,amsmath,amssymb,amsthm,amscd} 
\usepackage{float}
\usepackage{array,mathrsfs}
\usepackage{caption}
\usepackage{fancybox}\usepackage{enumerate}
\newcommand{\bea}{\begin{eqnarray}\displaystyle}
\newcommand{\eea}{\end{eqnarray}}

\newlength{\arrow}
{\setlength{\fboxsep}{15pt}
\setlength{\mylength}{\linewidth}%
\addtolength{\mylength}{-2\fboxsep}%
\addtolength{\mylength}{-2\fboxrule}%
\Sbox
\minipage{\mylength}%
\setlength{\abovedisplayskip}{0pt}%
\setlength{\belowdisplayskip}{0pt}%
\equation}%
{\endequation\endminipage\endSbox
\[\fbox{\TheSbox}\]}

\newdimen\tableauside\tableauside=1ex
\newdimen\tableaurule\tableaurule=0.8pt
\newdimen\tableaustep
\def\phantomhrule#1{\hbox{\vbox to0pt{\hrule height\tableaurule width#1\vss}}}
\def\phantomvrule#1{\vbox{\hbox to0pt{\vrule width\tableaurule height#1\hss}}}
\def\sqr{\vbox{%
  \phantomhrule\tableaustep
  \hbox{\phantomvrule\tableaustep\kern\tableaustep\phantomvrule\tableaustep}%
  \hbox{\vbox{\phantomhrule\tableauside}\kern-\tableaurule}}}
\def\squares#1{\hbox{\count0=#1\noindent\loop\sqr
  \advance\count0 by-1 \ifnum\count0>0\repeat}}
\def\tableau#1{\vcenter{\offinterlineskip
  \tableaustep=\tableauside\advance\tableaustep by-\tableaurule
  \kern\normallineskip\hbox
    {\kern\normallineskip\vbox
      {\gettableau#1 0 }%
     \kern\normallineskip\kern\tableaurule}%
  \kern\normallineskip\kern\tableaurule}}
\def\gettableau#1{\ifnum#1=0\let\next=\null\else
\squares{#1}\let\next=\gettableau\fi\next}

\begin{document}
\title{Quantum Residues and the Geometry of Perturbative/Non-perturbative Relations}
\author{Can Koz\c{c}az}
\affiliation{Department of Physics, Bo\u{g}azi\c{c}i University, Istanbul, Turkey}
\affiliation{Feza G\"{u}rsey Center for Physics \& Mathematics, Bo\u{g}azi\c{c}i University, Istanbul}
\emailAdd{can.kozcaz@bogazici.edu.tr}
\abstract{We develop a systematic, Whitham-based construction of perturbative/non-perturbative (P/NP) relations for one-dimensional quantum-mechanical systems admitting a genus-one WKB period description, and derive a higher-genus formulation under explicit regularity assumptions. We select external deformations that preserve the independent quantum residues and lift them by varying the internal moduli so that the independent perturbative quantum periods remain fixed. The Riemann bilinear identity then determines the induced evolution of the non-perturbative periods through local residue pairings and the inverse period Jacobian. At genus one, this reduces to a relation between the flow of the non-perturbative quantum period and the energy derivative with respect to the perturbative action, providing a constructive connection between the Rayleigh–Schrödinger expansion and fluctuations around non-perturbative saddles. We recover known P/NP relations and derive extensions for explicitly coupling-dependent potentials, which we verify at the computed orders through independent perturbative and non-perturbative period calculations for quantum-corrected quartic and Jacobi elliptic examples.}

\maketitle

\section{Introduction}

Perturbative and non-perturbative contributions to quantum-mechanical observables are typically calculated by different methods. Perturbation theory describes fluctuations around a chosen minimum. Non-perturbative effects, including tunneling and contributions associated with complex saddles, require information beyond the local perturbative expansion around that minimum. Their calculation is generally more demanding. Nevertheless, these two calculations are not independent. Their connection was already visible in early large-order analyses, which provided examples of what is now understood through resurgence \cite{Ecalle:1981,DDP,Voros1983}\footnote{Excellent reviews including modern developments include \cite{MarinoBook1,MarinoBook2,Dorigoni:2014hea,Aniceto:2018bis,Dunne:2025mye}}. The large-order behavior of perturbation theory carries information about other saddles, providing a connection between late and early terms of respective expansions \cite{Bender:1969si,Bender:1973rz}. The consistent combination of perturbative and multi-instanton sectors provides a description of the spectrum that is not available from either contribution separately \cite{Zinn-Justin:2004vcw, Zinn-Justin:2004qzw, Jentschura:2010zza}. The question is then how much non-perturbative information can be obtained directly from the perturbative expansion.  

A more constructive, now known as the P/NP relation, linking the low-order perturbative coefficients and the low-order fluctuations around non-perturbative saddles was developed \cite{Alvarez:2000,Alvarez2004,Dunne:2013ada,Dunne:2014bca,Dunne:2016qix,Dunne:2016jsr} for several one-dimensional systems admitting a genus-one description of their WKB periods. In this formulation, the perturbative function $B(E,g)$, or more precisely its inverse $E(B,g)$, can be used to determine derivatives of the non-perturbative function $A(B,g)$. This approach is different from extracting instanton data from the factorial growth of the large orders of the perturbative series, in that coefficients in the two sectors are related to each other at similar orders.

The P/NP relation was subsequently generalized to include one-parameter families of genus-one potentials, including the deformed anharmonic oscillator and Jacobi elliptic potentials in \cite{Cavusoglu:2023bai, Cavusoglu:2024usn}. For generic deformation parameters, the relevant residue is non-zero, in contrast to the standard residue-free examples underlying the original P/NP relation. In addition, the generalized P/NP relation for these deformed potentials were argued to take the following generic form once the deformation of the potential is expressed in terms of the residue,
\begin{equation} \label{parametricPNP_simplest}
 \partial_B E(B,\widetilde{R},g) = \frac{\partial_g \mathcal{A}(B,\widetilde{R},g)}{\partial_g \widetilde{S}(\widetilde{R},g)},
\end{equation}
where $\mathcal{A}$ is the non-perturbative quantum period. In this equation, $\widetilde {S}\equiv S/g$ and $\widetilde{R}\equiv R/g$ are the scaled instanton action and the scaled residue, respectively. The transformation of genus-one P/NP relations between different perturbative and non-perturbative cycle pairs, including the dependence on saddle frequencies and bion or bounce actions, was studied in \cite{Misumi:2025ijd}.

Motivated by the generic form of the P/NP relation, in this work, we develop a new systematic Whitham-based method to construct the generalized P/NP relations for the genus-one deformed potentials. The construction also suggests an extension to higher-genus families, where the fixed-period lift must accommodate several independent perturbative actions \cite{upcoming}. We first construct an external vector field that preserves the independent quantum residues and lift it by adjusting the energy to keep the perturbative quantum period fixed. The Riemann bilinear identity then expresses the variation of the non-perturbative period along this flow in terms of the energy derivative with respect to the perturbative action, with a coefficient calculated from the local residue pairing in the Riemann bilinear identity. The resulting equation provides a constructive relation between the Rayleigh–Schr\"{o}dinger perturbative expansion and fluctuations around non-perturbative saddles. In our formulation, the P/NP relation governs the evolution of the non-perturbative quantum period along a particular Whitham flow that preserves the perturbative quantum period and the independent quantum residues. The Riemann bilinear identity was crucially used to relate the perturbative and non-perturbative quantum periods earlier in \cite{vanSpaendonck:2023znn,Basar:2017hpr}

The connection between P/NP relations and Whitham dynamics has an important precedent in the work of Gorsky and Milekhin \cite{Gorsky:2014lia}. They  related the original P/NP relation with a Whitham equation governing the non-perturbative quantum period at fixed perturbative action. Their construction employs the Riemann bilinear identity and a coupling-dependent Whitham time in the Nekrasov-Shatashvili setting. They describe the resulting Whitham dynamics as deformed rather than quantizing the Whitham system itself. Our contribution is a systematic residue-preserving derivation of the deformation direction and its dual-period equation. We apply this construction to the genus-one families studied in \cite{Cavusoglu:2024usn} and to their quantum-corrected extensions. When the variation of the coupling alone preserves the quantum residues, no additional residue-compensating deformation is required. Otherwise, the residue conditions determine the necessary combined direction in parameter space.  

Although our explicit examples concentrate on genus-one, we also derive the higher-genus form of the construction under regularity assumptions, such as smoothness of the WKB curve, having enough internal moduli to keep the periods fixed and an invertible Jacobian for the independent perturbative quantum periods. In this case, preserving the independent perturbative quantum periods generally requires varying several internal moduli, rather than adjusting the energy alone. Provided these compensating variations preserve the quantum residues and the associated period Jacobian is invertible, the fixed-period lift is determined locally. The higher-genus Riemann bilinear identity then yields a coupled system governing the non-perturbative quantum periods, with coefficients expressed through local residue pairings and the inverse period Jacobian. This formulation reduces to our scalar relation at genus-one and separates the general geometric construction from the evaluation of the residue pairings for a particular potential. Related higher-genus P/NP relations have recently been studied in \cite{Ture:2026ktb}.

The paper is organized as follows. We first present our prescription for constructing P/NP relations for genus-one potential families. We briefly overview how our approach can be generalized to a higher-genus formulation locally on smooth WKB curve families and give the P/NP relation for an arbitrary sextic potential. In this paper, we mainly focus on genus-one potentials. We apply our construction to the potentials studied in \cite{Cavusoglu:2024usn} and to quantum-corrected quartic and Jacobi elliptic potentials. In the elliptic examples, we demonstrate cancellations between the quantum contributions of the original poles and the turning points to the bilinear residue pairing, through the WKB orders that we computed. We next derive the relation between the variation of the instanton action of the full corrected potential along the residue-preserving flow and the algebraic bilinear residue pairing. This relation makes explicit the role of the local oscillator frequency and clarifies how the action-based formulation in Eq.~\eqref{parametricPNP_simplest} extends to quantum-corrected potentials.  Some technical derivations and computational details are collected in the appendices.


\section{P/NP Relation as a Flow Equation}

\par{Different generalized forms of the original P/NP relation have been constructed for potentials with deformation parameters. These generalizations are distinguished by the non-vanishing residue of the WKB-differential, and it is argued that once the quantum residue $\mathcal{R}_q$ is treated as the natural deformation parameter instead of the arbitrary ones we choose (e.g. coefficients in front of the monomials for polynomial potentials or elliptic parameter for Jacobi elliptic potentials), the known generalized P/NP relation takes a remarkably simple and generic form. In this section, we would like to propose a general prescription for $deriving$ all of these known relations and others in a systematic way. Our construction consists of three steps. }
\begin{enumerate}[I.]
\item{\textit{Peripheral-cycle shift}}\\\\
The first step is to determine the vector field $\mathcal{V}$ that keeps the quantum residue $\mathcal{R}_q$ invariant, $\mathcal{V}\mathcal{R}_q=0$. We define the quantum residue by 
\begin{align}
\mathcal{R}_q= \operatorname{Res}_P\,\lambda,\qquad \text{with}\,\lambda\equiv \frac{1}{g}\, P\, dz,
\end{align}
where $P$ may lie above infinity (as is the case for polynomial potentials) or be a finite pole of the differential (as is the case for elliptic potentials). It is the branch-antisymmetric formal WKB momentum. For explicitly $g$-dependent potentials, its construction uses the corresponding modified Riccati recursion and is not equivalent to retaining only even powers of $g$ (c.f. Appendix \ref{appendixc}). Hence, we denote by $\lambda$ the full quantum corrected WKB differential. 

In this work, we consider families parametrized by $(E,g,\eta)$, with one shape-deformation parameter $\eta$, whose smooth WKB curves have genus one and whose residues are determined by an energy independent quantum residue $\mathcal{R}_q$,
\begin{align}
\partial_E \mathcal{R}_q=0.
\end{align}
Away from the discriminant, all statements are local in parameter space. 

The motivation to define the vector field $\mathcal{V}$ this way is to ensure the invariance under the residue-induced cycle transformations described below. We consider residue-induced monodromies for which analytic continuation around a relevant component of the discriminant locus changes the non-perturbative cycle according to $\mathfrak{b}\rightarrow\mathfrak{b}+n\delta_P,\,(n\in\mathbb{Z})$, with the perturbative cycle unchanged\footnote{The original physical discussion in detail can be found in \cite{Cavusoglu:2024usn}. A geometric derivation, which does not rely on the specifics of the instanton action, is presented in the Appendix \ref{appendixA}.}. The quantum period shift follows from the cycle transformation itself,
\begin{align}
\mathcal{A}\rightarrow\mathcal{A}+2\pi n\mathcal{R}_q(g,\eta).
\end{align}

In \cite{Cavusoglu:2024usn}, for the deformed anharmonic potential it has been noted that this shift can be directly observed from the explicit form of the instanton action $S(\eta)$. The monodromy around the discriminant locus at $\eta=1$ changes the instanton action by the quantum residue,
\begin{align}
\frac{1}{g}S(\eta)\rightarrow \frac{1}{g}S(\eta)+2\pi n \,\mathcal{R}_q(g,\eta),\qquad \mbox{for}\,n\in\mathbb{Z}.
\end{align}
due to the logarithmic branch cut. Thus, although the non-perturbative quantum period receives further WKB corrections, its shift under the stated peripheral-cycle transformation coincides with that of its leading classical term $(S/g)$ in this particular case (because the quantum residue does \textit{not} get any derivative WKB contributions for the polynomial potentials). However, we emphasize that the shift is captured by the shift of the instanton action is not universal.

The non-perturbative quantum period depends on the chosen representative of its cycle in the punctured curve. We require the proposed P/NP relation to be insensitive to the addition of a peripheral cycle, which shifts $\mathcal{A}$ by multiples of quantum residue. We argue that the P/NP relation can be understood as a particular Whitham flow of the non-perturbative quantum period $\mathcal{A}$, we therefore require that the residue must remain invariant under the flow,
\begin{align}
\mathcal{V}\mathcal{R}_q=0.
\end{align}  

\item{\textit{Fixed perturbative action}}\\\\
In general, the spectrum receives both perturbative and non-perturbative contributions. Quantization conditions can be recursively solved to determine the energy levels once the perturbative and non-perturbative quantum periods are known. The perturbative period is essentially the reduced action calculated in the classically allowed region and quantum corrections are captured by the WKB approximation. We generally denote the dimensionless perturbative quantum action by $B(E;g,\eta)$ \cite{Zinn-Justin:2004vcw, Zinn-Justin:2004qzw, Jentschura:2010zza},
\begin{align}
B(E;g,\eta)=\frac{1}{2\pi}\oint_{\mathfrak{a}}\lambda
\end{align}

Here $\mathfrak{a}$ encircles the branch cut joining the two turning points that bound the classically allowed region of the chosen well. If we are only interested in perturbative energy shifts for the $N$th level associated with the chosen perturbative well, we first regard $B$ as a continuous action coordinate when deriving the differential relation and then impose the following perturbative quantization condition,
\begin{align}
B=N+\frac{1}{2}.
\end{align}
This condition concerns the perturbative quantization around the chosen well, not the complete non-perturbative quantization condition. As we vary the deformation parameters of the potential, the perturbative quantum period varies as well and the energy changes accordingly, while keeping the quantum number $N$ fixed. In general, the perturbative period is \textit{not} invariant under the flow generated by $\mathcal{V}$ that keeps the quantum residue fixed: $\mathcal{V}B\neq0$. Therefore, we need to lift the vector field $\mathcal{V}$ by allowing the internal energy parameter $E$ to vary so that the perturbative period remains fixed:
\begin{align}
\widehat{\mathcal{V}}\equiv\mathcal{V}-\frac{\mathcal{V}B}{\partial_E B}\partial_E, 
\end{align}
where we further assume that $\partial_E B\neq0$. By construction, we have 
\begin{align}
\widehat{\mathcal{V}}\mathcal{R}_q=0, \qquad\mbox{and}\qquad\widehat{\mathcal{V}}B=0.
\end{align}

\item{\textit{Riemann bilinear identity}}\\\\
We claim that the P/NP relation is the flow of the non-perturbative quantum period $\mathcal{A}$ under the  residue and perturbative period preserving flow. The non-perturbative quantum period $\mathcal{A}$ is defined by integration over the dual cycle
\begin{align}
\mathcal{A}=\frac{1}{i}\oint_{\mathfrak{b}}\lambda.
\end{align}

The dual cycle $\mathfrak{b}$ connects the adjacent turning points of neighbouring wells across the classically forbidden barrier, passing along one sheet of the spectral curve and returning along the other. We introduce two one-forms,
\begin{align}
\omega_E\equiv \partial_E \lambda, \qquad\mbox{and}\qquad\Omega\equiv \widehat{\mathcal{V}}\lambda.
\end{align}
 By construction, we have
\begin{align}
\operatorname{Res}_{P}\,\Omega=0, \qquad\text{at every relevant pole }P,\qquad \oint_{\mathfrak{a}}\Omega=0.
\end{align}
The differential $\Omega$ is the variation of $\lambda$ normalized to have vanishing perturbative period. Its singular parts are determined by the chosen deformation and the WKB differential. These will be evaluated explicitly in the examples. With these singular data specified, $\Omega$ provides the quantum deformation of the corresponding normalized Whitham differential. 
Similar to \cite{Gorsky:2014lia}, we can now make use of the Riemann bilinear identity,
\begin{align}
\left(\oint_{\mathfrak{a}} \omega_{E}\right)\left(\oint_{\mathfrak{b}}\Omega \right)-\underbrace{\left(\oint_{\mathfrak{a}}\Omega \right)\left(\oint_{\mathfrak{b}}\omega_{E} \right)}_{=0}=-2\pi i \sum_{P}\operatorname{Res}_{P}\left( F_E\Omega\right),
\end{align}
where the second term vanishes by the Whitham flow condition we imposed on the perturbative period. The sign convention follows from our choice of the intersection of the cycles as $\mathfrak{a}\cdot\mathfrak{b}=-1$. We have defined the local primitive $dF_E=\omega_E$. The bilinear identity reduces to
\begin{align}
\big( \partial_E B\big)\big(\widehat{\mathcal{V}}\mathcal{A}\big)=-\mathcal{I}_{\widehat{\mathcal{V}}},
\end{align}
where we define the bilinear residue pairing
\begin{align}
\mathcal{I}_{\widehat{\mathcal{V}}}\equiv \sum_P\operatorname{Res}_P\left(F_E\Omega\right).
\end{align}
\end{enumerate}

Finally, we obtain the generic P/NP relation for genus-one potentials,
\begin{align}\label{PNP}
\boxed{\widehat{\mathcal{V}}\mathcal{A}=-\mathcal{I}_{\widehat{\mathcal{V}}}\frac{\partial E}{\partial B}}
\end{align}  


\section{Generalization to Higher Genus Potentials}

In this section, we develop a Whitham based construction to determine P/NP relations for genus one potentials in detail. However, our approach relies on principles which are not restricted to genus-one potentials. We can easily employ our construction to potentials whose WKB periods can be realized on higher genera curves. We will start with a polynomial potential that depends on a number of deformation parameters, which we will choose for now to be the coefficients. The residue is generically non-vanishing. Just as in the case of a genus-one potential, we find a vector field $\mathcal{V}$ that keeps the residue constant. In upcoming work \cite{upcoming}, we argue that the choice of $\mathcal{V}$ is less restrictive, and that the flow may be chosen to run in more than one direction. This freedom will allow us to construct other P/NP-like flow equations.
 
Following our prescription, we lift the vector field $\mathcal{V}$ to $\widehat{\mathcal{V}}$ to keep all the perturbative quantum periods $B_i$, $(i=1,\mathellipsis,g)$ invariant. Generically, the are $g$ holomorphic directions to lift $\mathcal{V}$, which perfectly matches the number of perturbative periods we would like to keep invariant. By construction, we again obtain a flow that satisfies
\begin{align}
\widehat{\mathcal{V}}\mathcal{R}_q=0, \qquad \text{and}\qquad \widehat{\mathcal{V}}B_i=0, \,\,\text{for}\,\,i=1,\mathellipsis g.
\end{align}

It will prove to be useful to introduce the quantum holomorphic-period Jacobian $\mathcal{J}$ by
\begin{align}
\mathcal{J}_{i\alpha}\equiv \partial_{a_\alpha}B_i,
\end{align}
and the one-forms $\Omega\equiv\widehat{\mathcal{V}}\lambda$ and $\omega_{\alpha}\equiv\partial_{a_\alpha}\lambda$, derivative along the holomorphic directions. As the final step, we can once more make use of the Riemann bilinear identity to get the flow equation for the non-perturbative quantum period:
\begin{align}
\boxed{\mathcal{J}^T\mathcal{W}=-\mathcal{I},}
\end{align}
where $\Phi$ is the local primitive of $\Omega$, satisfying $d\Phi=\Omega$. We call the bilinear pairing residue simply $\mathcal{I}$, which is defined by
\begin{align}
\mathcal{I}_{\alpha}\equiv \sum_{P}\operatorname{Res}_P\left(\Phi\omega_\alpha\right)
\end{align}
We get a matrix equation describing the flow for each of $g$ non-perturbative quantum periods. Above we introduced $\mathcal{W}$ for the notational simplicity, which is defined by
\begin{align}
i\mathcal{W}_i\equiv\oint_{\mathfrak{b}_i}\Omega, \qquad\text{for}\,\,i=1,\mathellipsis,g.
\end{align}
As an example, let us find the P/NP flow equations for a generic sextic potential given by,
\begin{align}
V(z)=z^6+a_4 z^4+a_3 z^3+a_2 z^2+a_1 z+a_0.
\end{align}
The associated flow equation (in the normalization we have been employing) becomes 
\begin{align}
\boxed{\widehat{\mathcal{V}}\mathcal{A}_i= \frac{1}{8g^2}\left[ (a_4-2)\frac{\partial a_0}{\partial B_i}+3a_3 \frac{\partial a_1}{\partial B_i}\right],\qquad i=1,2}
\end{align} 
Explicit checks of these equations will be provided in \cite{upcoming}.


\section{P/NP Relations for Deformed Potential}

In this section, we consider the examples studied in \cite{Cavusoglu:2024usn} and two new potentials employing our current method. The examples are chosen to demonstrate different aspects of genus-one potentials and the corresponding P/NP relation. 


\subsection{Deformed Anharmonic Oscillator}
\par{We start with the deformed anharmonic oscillator, whose WKB differential has a singularity at infinity. In what follows, we recover known forms of the P/NP relations from the generic form of the P/NP relation in Eq. (\ref{PNP}). The potential of the deformed anharmonic oscillator is given by }
\begin{align}
V(z;\eta)=\frac{1}{2}z^2-\eta z^3+\frac{1}{2}z^4,
\end{align}
where the deformation parameter is chosen to be the coefficient $\eta$ in front of the cubic monomial. The pole of the WKB differential is at infinity, and its quantum residue does not receive any derivative quantum corrections from the WKB expansion. The quantum residue is given by 
\begin{align}
\mathcal{R}_q(g,\eta)=\frac{i}{2g}\eta(\eta^2-1).
\end{align}
The $g$ in front is by the definition of the WKB differential (in \cite{Cavusoglu:2024usn}, they are called scaled quantum residue).According to our construction, we first need to find the vector field $\mathcal{V}$ keeping the residue invariant. The constant residue defines a curve in the $(g,\eta)$-plane, and the vector field $\mathcal{V}$ is tangent to this curve, but the speed of the flow is undetermined by this condition. We choose $\log g$ as a local parameter along the constant-$\mathcal{R}_q$ curves, imposing $\mathcal{V}g=g$. This normalization is natural from the WKB viewpoint and fixes the parametrization. Moreover, it will reproduce the known generalized P/NP relation for the other deformed genus-one potentials as well. Along the path with $\partial_\eta\mathcal{R}_q\neq 0$, the vector field with our parametrization is uniquely fixed,
\begin{align}\label{vectorfield}
\mathcal{V}=g\partial_g-g\frac{\partial_g \mathcal{R}_q}{\partial_{\eta}\mathcal{R}_q}\partial_{\eta}.
\end{align}
For the deformed anharmonic oscillator, the vector field $\mathcal{V}$ keeping the residue invariant along the flow is given by
\begin{align}
\mathcal{V}=g\partial_g+\frac{\eta(\eta^2-1)}{3\eta^2-1}\partial_\eta,
\end{align}
We can equivalently choose as the unnormalized characteristic field 
\begin{align}
\widetilde{\mathcal{V}}=(3\eta^2-1)g\partial_g+\eta(\eta^2-1)\partial_\eta.
\end{align}
Now let us lift the vector field $\mathcal{V}$ to the full parameter space $(E,g,\eta)$,
\begin{align}
\widehat{\mathcal{V}}=g\partial_g+\frac{\eta(\eta^2-1)}{3\eta^2-1}\partial_\eta+\dot{E}\partial_E .
\end{align}
where we have used the shorthand $\dot{E}$ for
\begin{align}\label{Edot}
\dot{E}=\widehat{\mathcal{V}}E=-\frac{g\partial_g B+\frac{\eta(\eta^2-1)}{3\eta^2-1}\partial_\eta B }{\partial_E B}.
\end{align}
By construction, the lifted vector field $\widehat{\mathcal{V}}$ leaves the perturbative quantum period $B$ invariant along the flow, as well as the quantum residue since it does not depend on the energy $E$. Having constructed the lifted vector field $\widehat{\mathcal{V}}$ to study the flow of the non-perturbative period $\mathcal{A}$, we are ready to implement the third step of our construction; determining the bilinear residue pairing to be able to make use of the Riemann bilinear identity. The WKB curve is given by the equation 
\begin{align}
p^2=2\xi-z^2+2\eta z^3-z^4,
\end{align}
where $\xi\equiv gE$. On the positive sheet near the infinity, the momentum behaves as
\begin{align}
p=iz^2-i\eta z+\frac{i(1-\eta^2)}{2}+\frac{i\eta(\eta^2-1)}{2z}+\mathcal{O}(z^{-2}).
\end{align}
We can easily compute $\omega_E$
\begin{align}
\omega_E=\frac{dz}{p}+\mbox{(terms that can not affect the residue)},
\end{align}
so the primitive $F_E$ can be written as 
\begin{align}
F_E=\frac{i}{z}+\frac{i\eta}{2z^2}+\frac{i(3\eta^2-1)}{6z^3}+\mathcal{O}(z^{-4}).
\end{align}
Finally, we can also find $\Omega$ to be
\begin{align}
\Omega=\frac{i}{g}\left[-z^2+\frac{2\eta^3}{3\eta^2-1}z+\frac{(\eta^2-1)^2}{2(3\eta^2-1)}+0\cdot \frac{1}{z}+\mathcal{O}(z^{-2}) \right].
\end{align}
Given $F_E$ and $\Omega$, we  compute $\mathcal{I}_{\widehat{\mathcal{V}}}$ by finding the appropriate residues
\begin{align}
\operatorname{Res}_{+\infty}(F_E\Omega)=\frac{1}{3g(3\eta^2-1)},
\end{align}
The residue is the same on the other sheet, since both factors in the residue change sign by going to the other sheet leaving their product the same. Therefore, we get
\begin{align}
\mathcal{I}_{\widehat{\mathcal{V}}}=\sum_{P=\pm\infty}\operatorname{Res}_{P}(F_E\Omega)=\frac{2}{3g(3\eta^2-1)}.
\end{align}
The dual flow equation takes the following form, which may look different than the generalized P/NP relation in \cite{Cavusoglu:2024usn},
\begin{align}
\boxed{\frac{\partial E}{\partial B}=-\frac{3}{2}g(3\eta^2-1)\widehat{\mathcal{V}}\mathcal{A}}
\end{align}
We can express the lifted vector field $\widehat{\mathcal{V}}$ in the coordinates $(B,g,\eta)$ (with $B$ being fixed) for this particular potential as
\begin{align}
\widehat{\mathcal{V}}=\left.g\frac{\partial}{\partial g}\right|_{B,\eta}+\frac{\eta(\eta^2-1)}{3\eta^2-1}\left.\frac{\partial}{\partial \eta}\right|_{B,g},
\end{align}
which leads to the P/NP relation for the nonperturbative quantum period,
\begin{align}
\boxed{\frac{\partial E}{\partial B}=-\frac{3}{2}g\left[(3\eta^2-1)g\frac{\partial\mathcal{A}}{\partial g} +\eta(\eta^2-1)\frac{\partial\mathcal{A}}{\partial \eta}\right ].}
\end{align}
Expressing $\mathcal{A}$ in term of the instanton function $A$, one recovers the generalized P/NP relation in Eq. (17) of \cite{Cavusoglu:2024usn}, For details, see Appendix \ref{appendixb}. 

Let us look more closely at the P/NP relation, and in particular the instanton action as a function of the deformation parameter $\eta$ of the potential 
\begin{align}
S(\eta)=-\frac{2}{3}+\eta^2+\frac{1}{2}\eta(\eta^2-1)\log\left(\frac{\eta-1}{\eta+1} \right).
\end{align}
The instanton action does not depend on the energy $E$, and if we define the scaled instanton action $\mathcal{S}\equiv S/g$, just like in \cite{Cavusoglu:2024usn}, we observe that
\begin{align}\label{instantonflow}
\mathcal{I}_{\widehat{\mathcal{V}}} =-1\cdot\widehat{\mathcal{V}}\mathcal{S}.
\end{align}
Let us make a few remarks on this equation\footnote{Notice that the factor stemming from the frequency of the harmonic vacuum is set to one by the choice of the potential. This factor will be discussed in \ref{instantonactionflow}.}. The instanton action $S$ is known to be part of the non-perturbative quantum period $\mathcal{A}$, which can be related to the instanton function $A$. It may be surprising that the residue in the Riemann bilinear identity might be related to the instanton action, but this equation suggests that the variation of the instanton action along the Whitham flow (as we keep the perturbative quantum period and the residue of the differential $\lambda$ fixed along this flow) is encoded by the sum of the residues $\mathcal{I}_{\widehat{\mathcal{V}}}$. We would like to state this curious relation as an observation rather than offering a physical explanation for the moment. 
The dual flow equation can now be rewritten as 
\begin{align}
\frac{\partial E}{\partial B}=\frac{\widehat{\mathcal{V}}\mathcal{A}}{\widehat{\mathcal{V}}\mathcal{S}},
\end{align}
which has precisely the structural form of Eq. (2) in \cite{Cavusoglu:2024usn}, but now obtained from the dual flow of the Whitham vector field.

\subsection{Jacobi SD Elliptic Potential}
\par{In the previous example, the residue of the WKB differential did not receive any quantum correction and its $g$ dependence was simple. Although this is a general feature of polynomial potentials, it is not generally true for potentials with poles at finite distances, like Jacobi elliptic potentials. Fortunately, our approach is capable of handling these types of potentials without ad hoc modifications. We focus on the Jacobi elliptic SD-square potential,}
\begin{align}
V(z;m)=\frac{1}{2}\text{sd}^2(z;m),
\end{align}
where $m$ is known as the elliptic parameter or modulus squared. We will treat it as the deformation parameter $\eta$ in our discussion. 

There is another new feature arising for the Jabobi elliptic potentials, the location of the pole $z_{*}$ varies as a function of the elliptic parameter $m$:
\begin{align}
z_{*}=K(m)+iK(1-m),
\end{align}
where $K(m)$ denotes the complete elliptic integral of the first kind. Before evaluating the quantum residue $\mathcal{R}_q(m)$, let us start with the classical residue $R(m)$ of the WKB differential with respect to the classical momentum $p_0$, satisfying $p_0^2=2\xi-\text{sd}^2(z,m)$, and 
\begin{align} \label{ellipticresidue}
R(m)=\frac{1}{\sqrt{m(1-m)}}.
\end{align}
We can use the following relation (derived in Appendix \ref{appendixc}, c.f. Eq. (\ref{riccattifull})) between the quantum corrected momentum $P$ and the classical momentum
\begin{align}\label{recursion}
P^2+\frac{g^2}{2}\frac{P''}{P}-\frac{3g^2}{4}\left( \frac{P'}{P}\right)^2=p_0^2,
\end{align}
to compute the quantum residue
\begin{align}
\mathcal{R}_q(m)=\sqrt{\frac{R(m)^2}{g^2}-\frac{1}{4}}.
\end{align}
Unlike in the case of polynomial potentials, the quantum corrections due the WKB expansion do contribute to the quantum residue at every order. 

We use again the particular parametrization for the vector field $\mathcal{V}$  in Eq. (\ref{vectorfield}),
\begin{align}
\mathcal{V}=g\partial_g-\frac{m(1-m)}{\frac{1}{2}-m}\partial_m.
\end{align}

Let us note that the quantum residue for the elliptic potential falls into a following particular class of quantum residues, which is encountered quite often in WKB computations,
\begin{align}
\mathcal{R}_q(\eta)=F\left( \frac{R(\eta)}{g}\right).
\end{align} 
for some function $F$. Once we look at the action of $\mathcal{V}$ on $\mathcal{R}_q(\eta)$ in this particular form, we get
\begin{align}
\mathcal{V}\mathcal{R}_q(g,\eta)=F'\left( \frac{R(\eta)}{g}\right)\,\mathcal{V}\left( \frac{R(\eta)}{g}\right)=0.
\end{align}
Therefore, even though the quantum residue itself receives genuine corrections, as long as it has this functional dependence it suffices that $\mathcal{V}$ keeps the ratio $R(\eta)/g$ invariant, where $R(\eta)$ is the residue computed for the classical momentum $p_0$. This is a useful and slightly surprising point; although the  quantum correction to the residue at finite pole changes the invariant residue value, the residue-preserving direction does not, as long as it is has the form $F(R(\eta)/g)$. 

As the second step, we lift the vector field $\mathcal{V}$ to $\widehat{\mathcal{V}}$ keeping the perturbative quantum period invariant along the flow. We can then construct the P/NP relation after evaluating the bilinear residue pairing. For the residue calculation, let us make the following coordinate change:
\begin{align}\label{change}
u=\text{sn}(z;m),
\end{align}
which allows us to write the WKB curve as
\begin{align}
y^2=(1-u^2)(2\xi(1-mu^2)-u^2),
\end{align}
where we have defined $y=P\frac{du}{dz}$. In the new coordinates $(u,y)$, the leading WKB differential takes the following form
\begin{align}
\lambda_0=\frac{2\xi(1-mu^2)-u^2}{g(1-mu^2)y}du.
\end{align}
Note that we are reparametrizing, not modifying the potentials or the periods. The next step is to identify the one-forms that enter in the definition of the residue, for simplicity, we consider the  classical and the quantum computations separately.
\subsubsection{Classical Contributions}
 We start with the classical part of the aforementioned one-forms and then check for possible quantum corrections:
\begin{align}
\omega_E^{(0)}=\partial_E\lambda_0=\frac{du}{y},
\end{align}
and 
\begin{align}
\Omega^{(0)}=\left[E-\frac{\mathcal{V}B}{B_E}-\frac{2gE(1-mu^2)-u^2}{g(1-mu^2)}+ \frac{\beta u^2(gE(1-mu^2)-u^2)}{g(1-mu^2)^2}\right]\frac{du}{y},
\end{align}
where we have defined $\beta=R(m)/R'(m)$. In $(u,y)$ coordinates, the poles of $\lambda_0$ occur at 
\begin{align}
u_{\pm}=\pm\frac{1}{\sqrt{m}}.
\end{align}
Above each value of $u$, there are two points of the curve 
\begin{align}
y_{*}=\pm\frac{\sqrt{1-m}}{m}.
\end{align} 
Thus this algebraic representation contains four pole points. Their residues are all determined by the same single independent residue datum. In this case, it turns out all four (classical) residues are the same,
\begin{align}
\mathcal{I}^{(0)}_{\widehat{\mathcal{V}}} =4\operatorname{Res}_{P_{*}}\left(F_E^{(0)}\Omega^{(0)} \right)=\frac{1}{g(\frac{1}{2}-m)}
\end{align}

\subsubsection{Quantum Corrections}

Now we turn our attention to the quantum corrections by WKB. Unlike the polynomial potential we studied before, at the quantum level, we have two distinct types of quantum corrections. One set of corrections are associated with the pole of the potential and another one stems from the turning points. The contributions from both of these sources are  individually non-vanishing, but they cancel each other when we put them together to evaluate the bilinear pairing residue. 

Let us start with the quantum corrections to the bilinear residue pairing at the pole of the potential. We only need to expand the quantum corrected momentum up to the linear order around the pole $z_{*}$ of the Jacobi SD-function since no higher terms in this expansion can generate a pole to contribute to the residue,
\begin{align}
p=\frac{g\mathcal{R}_q}{w}+b_1 w+\mathcal{O}(w^3), \qquad w\equiv z-z_{*}.
\end{align}
We can find the linear coefficient $b_1$ by focusing on the constant term of Eq. (\ref{recursion}) to be
\begin{align}
b_1=\frac{\mathcal{R}_q\left[2gE+R(m)^2(1-2m)/3\right]}{2g (\mathcal{R}_q^2+1)}.
\end{align}
The dependence of $b_1$ on the quantum residue $\mathcal{R}_q$ is exact, which means that the coefficients includes all quantum corrections summed up. This allows us to compute the one-form $\omega_E$,
\begin{align}
\omega_E=\frac{\mathcal{R}_q}{g(\mathcal{R}_q^2+1)}w\,dw+\mathcal{O}(w^3)\,dw.
\end{align}
We can also choose the alternative local coordinate $u$ that we previously used in Eq. \eqref{change},
\begin{align}
\omega_E=\frac{\mathcal{R}_q}{g(\mathcal{R}_q^2+1)(1-m)u_{*}}dt+\mathcal{O}(t)\,dt,\qquad t\equiv u-u_{*},
\end{align}
with the appropriate sign of $\mathcal{R}_q$ at the selected pole. The deformation one-form $\Omega$ along the flow is given by
\begin{align}
\Omega=\frac{\mathcal{R}_q}{2}\left(\mathcal{V}u_{*}(m)\right) \frac{dt}{t^2}+\mathcal{O}(1)dt.
\end{align}
Note that the pole is expressed as a function of the elliptic parameter, and that the vector field $\mathcal{V}$ includes a derivative with respect to this deformation parameter. 

The sum of the original four poles leads to the following \textit{exact} result:
\begin{align}
\mathcal{I}^{\rm poles}_{\widehat{\mathcal{V}}} &=\frac{1}{g\left(\frac{1}{2}-m\right)}\frac{\mathcal{R}_q^2}{\mathcal{R}_q^2+1},\\
&=\frac{1}{g\left(\frac{1}{2}-m\right)}\frac{1-\frac{1}{4}g^2m(1-m)}{1+\frac{3}{4}g^2m(1-m)}.
\end{align}
We can expand this residue (we will check the turning point contributions up to order $\mathcal{O}(g^4)$),
\begin{align}
\mathcal{I}^{\rm poles}_{\widehat{\mathcal{V}}} =\frac{1}{g\left(\frac{1}{2}-m\right)}\left[ 1-g^2m(1-m)+\frac{3}{4}g^4m^2(1-m)^2+\mathcal{O}(g^6)\right].
\end{align}

The last expression may be somewhat concerning at first sight, given the P/NP relationship satisfied by the same potential in \cite{Cavusoglu:2024usn}. The classical contribution here is interpreted as the coefficient there. The apparent discrepancy arises because we have not fully accounted for all contributions to the bilinear pairing residue yet, as we mentioned earlier. If we consider the quantum corrections by WKB to the pairing for the polynomial potentials, they do not contribute since the degree of the pole is increasing with degree of the quantum corrections. It turns out that this is not the case for the elliptic potential that we are considering here, and we get further contributions to $\mathcal{I}_{\widehat{\mathcal{V}}}$. As we will explicitly show for the first two corrections, they precisely cancel the quantum corrections that we calculated above. We conjecture, based on our results and those of \cite{Cavusoglu:2024usn}, that all quantum corrections coming from the turning point poles perfectly cancel the quantum corrections at the singularity of the WKB differential. In other words, we only see the classical part of the bilinear pairing residue explicitly in the P/NP relation not due to the lack of quantum corrections, but because there are intricate cancellations between two distinct types of quantum corrections. We will leave the proof of these cancellations for future work. 

Let us briefly sketch the calculation of the turning point contributions. We will introduce some notation to simplify the formulas:
\begin{align}
h(u)=1-mu^2,\qquad N(u)=2\xi h(u)-u^2\qquad v(u)=(1-u^2)h(u).
\end{align}
Let us write the quantum corrected WKB momentum $P$ and the WKB differential $\lambda$ as 
\begin{align}
P=p_0\left(1+g^2f_2+g^4f_4+\cdots\right),\qquad
\lambda=\sum_{n\geq0}g^{2n-1}f_{2n}\sqrt{\frac qv}\,du,\qquad f_0=1.
\end{align}
We will restrict ourselves to check the cancellations up to $\mathcal{O}(g^4)$, hence, we only need $f_2$ and $f_4$,
\begin{align}
f_2&=\frac{1}{32q^3}\left[5vq_u^2 -4q\left(vq_{uu}+\frac{1}{2}v_uq_u\right) \right],\\
f_4&=-\frac{1}{2}f_2^2-\frac{1}{4q}\left[vf_{2,uu} +\left(\frac{1}{2}v_u-\frac{vq_u}{2q} \right)f_{2,u}\right],
\end{align}
where we used $q=p_0^2$. The one-form $\omega_E=\partial_E \lambda$, which goes into the residue has an expansion in $g$ that descends from the expansion of $\lambda$
\begin{align}
\omega_E=\sum_{n\geq0}g^{2n}\frac{\mathscr{E}_{2n}}{y}du,\qquad\mathscr{E}_{2n}\equiv f_{2n}+q\,\partial_\xi f_{2n},
\end{align}
with $\partial_E=g\partial_\xi$. For simplicity, we use the regular residue preserving vector field $\mathcal{W}$ instead of $\mathcal{V}$, which satisfies:
\begin{align}
\mathcal{W}=\left(\frac{1}{2}-m\right)\mathcal{V}=\left(\frac{1}{2}-m\right)g\partial_g-m(1-m)\partial_m.
\end{align}
We can similarly expand the one-form $\mathcal{W}\lambda$ as
\begin{align}
\mathcal{W}\lambda=\sum_{n\geq0}g^{2n-1}\frac{\mathscr{W}_{2n}}{y}du,
\end{align}
where we can also explicitly write $\mathscr{W}_{2n}$ as 
\begin{align}
\mathscr{W}_{2n}=\left(\frac{1}{2}-m \right) \left[(2n-1)qf_{2n}+\xi\mathscr{E}_{2n} \right]-m(1-m)q\left[f_{2n,m}+\frac{1}{2}\left(\frac{q_m}{q}-\frac{v_m}{v}\right)f_{2n} \right].
\end{align}
Note that the flow variation of the WKB one-form is with respect to the lift of the vector field $\widehat{\mathcal{W}}$, but we can just use $\mathcal{W}$ since
\begin{align}
\operatorname{Res}\left[F_E\left(-\frac{\mathcal{W}}{\partial_E B}\omega_E\right) \right]=0, \qquad\text{with}\,\,F_E\omega_E=\frac{1}{2}d(F^2_E).
\end{align}
 The turning points $\sigma$ are the zeros of the function $N(u)$, satisfying
\begin{align}
\sigma^2=\frac{2\xi}{1+2m\xi}.
\end{align}
We expand $u$ around the turning point $\sigma$ to understand the singular behaviour of the pairing around the turning point. Setting $u=\sigma+t$, we can expand in $t$ 
\begin{align}
\frac{\mathscr{E}_{2n}}{y}&=\sqrt{-\frac{1-m\sigma^2}{2\sigma(1-\sigma^2)}}\sum_{k\geq0}a_{n,k}t^{k-3n-\frac{1}{2}},\\
\frac{\mathscr{W}_{2n}}{y}&=\sqrt{-\frac{1-m\sigma^2}{2\sigma(1-\sigma^2)}}\sum_{k\geq0}b_{n,k}t^{k-3n-\frac{1}{2}}
\end{align}
Using the first expansion we can construct the local primitive $F_E$ and then evaluate the bilinear pairing residue at order $g^{2N-1}$:
\begin{align}
\left[g^{2N-1}\right]\operatorname{Res}_{u=\sigma}\left( F_E\mathcal{W}\lambda\right)=-\frac{1-m\sigma^2}{\sigma(1-\sigma^2)}\sum_{\substack{i+j=N\\ k+\ell=3N-1}}\frac{a_{i,k}b_{j,\ell}}{k-3i+\frac{1}{2}},
\end{align}
which shows us that the quantum corrections to the residue are generically non-vanishing. The first two non-vanishing contributions are given by
\begin{align}
\mathcal{I}^{\rm tp}_{\widehat{\mathcal{V}}} =\frac{1}{g\left(\frac{1}{2}-m\right)}\left[ g^2m(1-m)-\frac{3}{4}g^4m^2(1-m)^2+\mathcal{O}(g^6)\right].
\end{align}
Once we combine all quantum corrections to the order that we computed, we get the advertised result,
\begin{align}
\mathcal{I}_{\widehat{\mathcal{V}}}=\mathcal{I}^{\rm poles}_{\widehat{\mathcal{V}}}+\mathcal{I}^{\rm tp}_{\widehat{\mathcal{V}}}=\frac{1}{g\left(\frac{1}{2}-m\right)}\left[1+\mathcal{O}(g^6)\right].
\end{align}
We therefore conclude that non-zero local quantum contributions cancel each other, i.e. only the classical piece appears after perfect cancellations, not due to the absence of quantum corrections.

\subsection{Quantum Corrected Polynomial Potentials}

We looked at potentials giving rise to genus-one WKB curves that did not involve any terms including $g$. Potentials involving terms with $g$ arise, for example, when we couple a number of Grassmann variables and project to fixed fermion number subsectors. The resurgence structure of such a system was studied in \cite{Behtash:2015loa,Kozcaz:2016wvy}, where the fermion number was allowed to take real values $\zeta$ to expose the hidden divergences in perturbation theory. Interestingly, the original form of the P/NP relation seemed to work for the potential of the form
\begin{align}
U=\frac{1}{2}\left( W'\right)^2+\frac{\zeta g}{2}W''.
\end{align}
In particular, for the tilted double well the ``superpotential'' is given by 
\begin{align}
W(z)=\frac{z^3}{3}-\frac{z}{4}.
\end{align} 
In this case, the potential $U$ becomes a quantum corrected potential with a linear term involving $g$
\begin{align}
U(z;g)= \frac{1}{2}\left(z^2-\frac{1}{4}\right)^2+g\zeta z.
\end{align}
In this section, we would like to use our flow construction to understand why the original P/NP relation worked for the above potential. Surprisingly, we find that the original P/NP relation requires modification when considering quantum corrections that are quadratic in $z$. We construct the relevant P/NP relation and check its validity by explicitly constructing the quantum periods using Picard-Fuchs equations. 

We want to keep our discussion somewhat more general and do not want to restrict ourselves to a quantum corrected potential arising from generalizations of the supersymmetric quantum mechanics with a number of Grassmann valued variables. Instead, we start our discussion with the following form of the potential, where we keep $V_{1}(z;\eta)$ independent of $V_{0}(z;\eta)$:
\begin{align}
U(z;g,\eta)=V_{0}(z;\eta)+gV_{1}(z;\eta),
\end{align}
where $V_{0}(z;\eta)$ determines the WKB curve, which we still take to be genus one. We will refer to it as the classical potential. In the presence of the quantum corrections to the potential, the exact WKB approach requires slight modification, discussed in Appendix \ref{appendixc}. 

First, we choose the quantum correction to the potential $V_{1}$ linear in $z$, just as in \cite{Kozcaz:2016wvy}:
\begin{align}
U(z;g)=\frac{1}{2}z^2(z-1)^2+g\alpha z,
\end{align}
where we will keep $\alpha$ fixed for simplicity. It is straightforward to compute the quantum corrected residue, 
\begin{align}
\mathcal{R}_q=-i\alpha.
\end{align}
Note that the quantum residue does \textit{not} depend on $g$, in contrast with the previous quantum residues of the WKB differential that we computed. Therefore, we do not need any correction term for the residue preserving vector field except for the flow with respect to $g$:
\begin{align}
\mathcal{V}=\left. g\partial_g\right|_{E,\alpha}.
\end{align}
We can lift to $\widehat{\mathcal{V}}$ the same way as we did before. The last step to write the P/NP relation requires the evaluation of the bilinear pairing residue at $\pm\infty$,
\begin{align}\label{pairingresiduelinear}
\mathcal{I}_{\widehat{\mathcal{V}}} =2\operatorname{Res}_{+\infty}\left(F_E\Omega \right)=\frac{1}{3g}
\end{align} 
For the quantum dual period $\mathcal{A}$, we obtain the following flow equation:
\begin{align}
\boxed{\frac{\partial E}{\partial B}=-3g^2\left.\frac{\partial\mathcal{A}}{\partial g}\right|_{B,\alpha}}
\end{align}
The dual quantum period $\mathcal{A}$ is related to the instanton function $A$ by
\begin{align}
\mathcal{A}=A+(2B-\alpha)\log\frac{g}{2}+C(B,\alpha),
\end{align}
leading to the more familiar form of the P/NP relation appearing in the literature,
\begin{align}
\frac{\partial E}{\partial B}=-3g\left[2B-\alpha+g\frac{\partial A}{\partial g} \right].
\end{align}
We recover the original P/NP relation with a simple shift in the energy.

Next, suppose we have the following quantum corrected potential,
\begin{align}
U(z;g)=\frac{1}{2}z^2(z-1)^2+gc z^2.
\end{align}
The quantum residue is once again constant and does not depend on $g$,
\begin{align}
\mathcal{R}_g=-ic.
\end{align}
The vector field $\mathcal{V}$ and its lift $\widehat{\mathcal{V}}$ do not change, but crucially the total bilinear pairing residue changes,
\begin{align}
\mathcal{I}_{\widehat{\mathcal{V}}} =\frac{1+2cg}{3g}.
\end{align} 
We obtain a modified P/NP relation for the dual quantum period $\mathcal{A}$ in this case:
\begin{align}
\boxed{\frac{\partial E}{\partial B}=-\frac{3g^2}{1+2cg}\left.\frac{\partial\mathcal{A}}{\partial g}\right|_{B,c}.}
\end{align}
We can write this relation more explicitly as 
\begin{align}
\frac{\partial E}{\partial B}=-\frac{3g}{1+2cg}\left(g\partial_g\mathcal{A}-g\frac{\partial_g B}{\partial_E B} \partial_E\mathcal{A}\right)
\end{align}
The non-perturbative instanton function $A(E,g,c)$ can be extracted from the non-perturbative quantum period after the propagation across the barrier and the local normalization contributions from the two wells are subtracted \cite{Kamata:2021jrs}. This can be written as 
\begin{align}
A(E,g,c)=\mathcal{A}-(2B-c)\log\frac{g}{2}-G(B)-G(B-c)+2c,
\end{align}
where $F(B)$ is defined by
\begin{align}
G(B)=\log\Gamma\left(B+\frac{1}{2} \right)-\frac{1}{2}\log(2\pi).
\end{align}
We can re-write the P/NP relation in terms of $A(E,g,c)$
\begin{align}
(1+2cg)\partial_B E+3g(2B-c+g\partial_g A)=0.
\end{align}
Note that this relation might appear singular due to the factor $1+2cg$, because it vanishes at $c=-1/2g$. However, this point should be excluded from our consideration since at this particular value of the coefficient $c$ in the quantum correction the potential ceases to have a harmonic minimum.

This particular quantum corrected potential was not studied in the context of the P/NP relations before, to the best of our knowledge. We checked this P/NP relation up to $\mathcal{O}(g^6)$ using the perturbative and non-perturbative quantum periods that we computed by solving the Picard-Fuchs equation\footnote{We thank E. O. Kaya for useful discussions.} and the quantum corrections acting by the appropriate operators described in \cite{Fischbach:2018yiu} to the classical periods. However, for completeness, here we provide the $A$ and $B$ functions only up to $\mathcal{O}(g^4)$ to save space,
\begin{align}\nonumber
B(E,g,c)&=E+g\Bigg[3E^2-cE+\frac{1}{4} \Bigg]+g^2\Bigg[35E^3-\frac{45}{2}cE^2+\left( \frac{3}{2}c^2+\frac{25}{4}\right)E-\frac{13}{8}c \Bigg]\\\nonumber
&+g^3\Bigg[ \frac{1155}{2}E^4-525cE^3+\left(105c^2+\frac{735}{4} \right)E^2-\left(\frac{5}{2}c^3+\frac{385}{4}c \right)E\\
&+\frac{25}{4}c^2+\frac{175}{32}\Bigg]+\mathcal{O}(g^4),
\end{align}
though higher orders are straightforward.
The non-perturbative instanton function $A(E,g,c)$ can be extracted from the non-perturbative quantum period
\begin{align}\nonumber
A(E,g,c)&=\frac{1}{3g}+g\Bigg[17E^2-5cE-\frac{1}{2}c^2 +\frac{19}{12}\Bigg]+g^2\Bigg[227E^3-\frac{269}{2}cE^2\\\nonumber
&+\left(10c^2+\frac{187}{4}\right)E+\frac{1}{12}c^3-\frac{87}{8}c\Bigg]+g^3\Bigg[ \frac{47431}{12}E^4-\frac{20395}{6}cE^3\\\nonumber
&+\left(\frac{2671}{4}c^2+\frac{34121}{24} \right)E^2-\left(\frac{115}{6}c^3+\frac{16613}{24}c \right)E-\frac{1}{24}c^4+\frac{2185}{48}c^2\\\nonumber
&+\frac{28829}{576}\Bigg]+\mathcal{O}(g^4).
\end{align}
These examples demonstrate that the residue preserving flow direction determined by $\mathcal{V}$ is by itself not enough. Second, the bilinear pairing plays an important role. We will demonstrate this in the last example. The local bilinear pairing probes more of the local Laurent data through $F_E\mathcal{V}\lambda$.

\subsection{Quantum Corrected Elliptic Potential}

We observed earlier that the potentials with poles at finite point exhibit subtle differences compared to the ones with poles at infinity. Hence, we would also like to introduce an elliptic quantum correction to the elliptic potential that we have studied earlier. We choose our potential from the familiar class of quantum-deformed potentials built from $W'^2$ and $gW''$, such that $U=\frac{1}{2}W'^2+gc W''$:
\begin{align}
U(z;g,m,c)=\frac{1}{2}\text{sd}^2(z;m)+gc\frac{\text{cn}(z;m)}{\text{dn}^2(z;m)}
\end{align}
This quantum correction changes the relative depths of the wells and their local frequencies, while the period problem still admits a genus-one algebraic representation. It also changes the finite-pole residues, so it tests the parts of our construction that are not probed by the polynomial examples.

Our construction starts with determining the residues at the poles of the potential. Our deformed potential has two inequivalent poles (note that all the known P/NP relations were constructed for potential with a single pole and associated residue so far, in this sense this quantum corrected potential is special), which the classical part of the potential cannot distinguish:
\begin{align}
z_1=K(m)+iK(1-m),\qquad z_2=3K(m)+iK(1-m).
\end{align}
We can evaluate the associated residues at these poles by expanding around them,
\begin{align}
U(z_1+w;g,m,c)&=\left(-\frac{R(m)^2}{2}+igcR(m) \right)\frac{1}{w^2}+\mathcal{O}(1),\\
U(z_2+w;g,m,c)&=\left(-\frac{R(m)^2}{2}-igcR(m) \right)\frac{1}{w^2}+\mathcal{O}(1),
\end{align}
where $R(m)$ is the classical residue mentioned in Eq. (\ref{ellipticresidue}). Next we compute the quantum residues including the derivative corrections due to WKB,
\begin{align}
\mathcal{R}_{q,\pm}=\sqrt{\left(\frac{R(m)}{g} \right)^2\pm 2ic\frac{R(m)}{g}-\frac{1}{4}},
\end{align}
where we distinguished the residue at the poles $z_{1,2}$ by $\pm$ because they differ by a sign. Each residue has an opposite-sheet partner. 

The Whitham flow is defined by requiring that both residues remain invariant under it,
\begin{align}
\mathcal{V}\mathcal{R}_{q,+}=0, \qquad \mathcal{V}\mathcal{R}_{q,-}=0.
\end{align}
This can be equivalently expressed as two independent conditions,
\begin{align}
\mathcal{V}(R(m)/g)=0\qquad \text{and}\qquad \mathcal{V}c=0. 
\end{align}
Initially, we did not need to assume anything about $c$, but since it can be expressed in terms of the quantum and classical residues as
\begin{align}
c=\frac{\mathcal{R}_{q,+}^2-\mathcal{R}_{q,-}^2}{4i(R/g)},
\end{align}
$c$ must be also invariant along the flow that keeps both quantum residues fixed. Let us choose again our logarithmic-coupling normalization $\mathcal{V}g=g$. The flow requires that the elliptic parameter $m$ changes along the flow as $\mathcal{V}m=R/R'$ such that we can write
\begin{align}
\mathcal{V}=g\partial_g-\frac{m(1-m)}{\frac{1}{2}-m}\partial_m.
\end{align}
The necessary lift of the vector field to keep the perturbative periods invariant does not spoil residue preserving conditions since the residues are independent of the energy $E$. The essential point is that the two pole types constrain different combinations of the parameters. Their common part fixes the scaled classical residue $R/g$, while their relative splitting fixes the deformation parameter $c$. This is why preserving the complete quantum-residue data selects the old elliptic direction without a $\partial_c$ component

The next step in our prescription is to determine the lifted vector field that additionally keeps the perturbative period fixed, 
\begin{align}
\widehat{\mathcal{V}}B=0.
\end{align}
Finally, we determine the residue of the bilinear pairing that comes from the Riemann bilinear relation. 

\subsubsection{Algebraic Contributions}
Similar to the case of the elliptic potential without any quantum correction to it, we should consider the contributions coming from quantum corrections by WKB to the residue at the poles separately from the contributions coming from the simple poles at the turning points. Let us perform a coordinate transformation to evaluate the residues,
\begin{align}
u=\text{cn}(z;m).
\end{align} 

For notational simplicity, let us introduce some functions,
\begin{align}
h(u)\equiv1-m+mu^2,\qquad N(u)=2\xi h(u)-1+u^2-2gcu.
\end{align}
We can rexpress the \textit{algebraic}\footnote{Previously, we would call the momentum classical momentum in the absence of any quantum correction to the potential, since it would not include $g$. However, the momentum depends on $g$ even before we include the derivative corrections.}  momentum and the associated curve to our potential,
\begin{align}
p^2_{\rm alg}=\frac{N(u)}{h(u)}, \qquad y^2=(1-u^2)N(u).
\end{align}
Including the Jacobian associated with the coordinate transformation, the WKB one-form and its derivative with respect to the energy are given by
\begin{align}
\lambda_{\rm alg}=\frac{N(u)}{g\,h(u)y}du, \qquad \partial_E\lambda_{\rm alg}=\frac{du}{y}
\end{align}
All parameter derivatives in this representation are taken at fixed $u$, including the explicit parameter dependence of the Jacobian.

The classical contribution to the perturbative period starts with the action of the particle oscillation around the minimum $z_0$ of the potential; if we denote the turning points by $z_{\pm}$, the half of the period can be described by $z_+\rightarrow z_0\rightarrow z_-$. The coordinate transformation from $z$ to $u$ maps both turning points $z_{\pm}$ to the same point $u_t$ and the minimum point $z_0$ to $-1$; the path we mentioned becomes $u_t\rightarrow -1\rightarrow u_t$. We need to keep track of the integration boundaries and the sign changes of the integrand when we perform the period integrals.
\begin{align}
\frac{1}{g}\int_{z_0}^{z_+}p_0\, dz=\int_{-1}^{u_t}\frac{\sqrt{N(u)}}{gh(u)\sqrt{1-u^2}}du.
\end{align}
We can similarly write for the integral from the left turning point to the minimum as 
\begin{align}
\frac{1}{g}\int_{z_-}^{z_0}p_0\, dz=-\int_{u_t}^{-1}\frac{\sqrt{N(u)}}{gh(u)\sqrt{1-u^2}}du,
\end{align}
where the minus sign originates from the fact that $du/dz=h(u)\sqrt{1-u^2}$ changes sign as $z$ passes $z_0$. Hence, we end up with a factor of $2$ once we define the closed path in the $u$-variable: 
\begin{align}
\frac{1}{g}\int_{z_-}^{z_+}p_0\, dz=2\int_{-1}^{u_t}\frac{\sqrt{N(u)}}{gh(u)\sqrt{1-u^2}}du.
\end{align}
The perturbative and non-perturbative quantum periods are defined accordingly as
\begin{align}
B=\frac{1}{\pi}\oint_{\mathfrak{a}}\lambda,\qquad\mathcal{A}=\frac{1}{2i}\oint_{\mathfrak{b}}\lambda.
\end{align}
The Riemann bilinear identity with these pre-factors in front of the integrals becomes,
\begin{align}
\left( \oint_{\mathfrak{a}}\omega_E\right)\left(\oint_{\mathfrak{b}}\Omega\right)=\big(\pi B_E \big)\big(2i\widehat{\mathcal{V}}\mathcal{A} \big)=-2\pi i\sum_P\operatorname{Res}_P(F_E\Omega).
\end{align}
The final step to get the P/NP relation is the computation of the bilinear pairing residue. We start with calculating the contributions from the algebraic part of the WKB differential; its poles are 
\begin{align}\label{poles}
u_{*}=\pm i\sqrt{\frac{1-m}{m}}.
\end{align}
The fixed branch points $u=\pm1$ do not contribute to the bilinear residue sum on a smooth fiber. Although $y$ vanishes there, the original momentum is nonzero, and the differentials are regular in the local coordinate $s=\sqrt{u\mp1}$. Their role changes only when a genuine turning point collides with them at $\xi=\pm gc$. The potential has two projected pole locations, given by the roots of $h(u)$. Since these roots are not branch points of the spectral curve, each has two distinct pre-images distinguished by the sign of $y$. The WKB differential consequently has four pole points on the spectral curve, with opposite residues at each sheet pair. Now, we can expand the differential around its poles by defining $t=u-u_{*}$
\begin{align}
\lambda_{\rm alg}=r_{*}\frac{dt}{t}+\mathcal{O}(1)dt, \qquad\text{with}\qquad r_{*}=\frac{N(u_{*})}{2gmu_{*}y_{*}}
\end{align}
Along the flow the residue is preserved by construction, but as we change the elliptic parameter $m$ the pole moves,
\begin{align}
\mathcal{V}r_{*}=0,\qquad\frac{\mathcal{V}u_{*}}{u_{*}}=-\frac{1}{1-2m}.
\end{align}
Using these identities, we can show that the one-form controlling the flow of the WKB differential $\Omega_{\rm alg}$ and the local primitive $F^{\rm alg}_E$ are given by
\begin{align}
\Omega_{\rm alg}=r_{*}(\mathcal{V}u_{*})\frac{dt}{t^2}+\mathcal{O}(1)dt, \qquad\text{and}\qquad F^{\rm alg}_E=\frac{t}{y_{*}}+\mathcal{O}(t^2),
\end{align}
which allows us to calculate the bilinear pairing residue at a single pole $P_{*}$,
\begin{align}
\operatorname{Res}_{P_{*}}\big(F^{\rm alg}_E \Omega_{\rm alg}\big)=\frac{1}{4g\left( \frac{1}{2}-m\right)}.
\end{align}
The signs of $F^{\rm alg}_E$ and $\Omega_{\rm alg}$ change at different poles such that their product has always the same sign, therefore we get 
\begin{align}
\mathcal{I}^{\rm alg}_{\widehat{\mathcal{V}}} =\frac{1}{g\left( \frac{1}{2}-m\right)}.
\end{align}
\subsubsection{Quantum Corrections}
We have deliberately avoided including the derivative WKB corrections so far. There are two different types of poles that can contribute the residue independently, and we need to treat them separately as we previously did. We avoid the details of the computations as they are essentially very similar to the ones we performed for the elliptic potential without any quantum correction before, and instead just state the results. The residue contributions can be calculated exactly as before,
\begin{align}\nonumber
\mathcal{I}^{\rm poles}_{\widehat{\mathcal{V}}}&=\frac{1}{g(1-2m)}\left[\frac{\mathcal{R}^{2}_{q,+}}{1+\mathcal{R}^{2}_{q,+}}+\frac{\mathcal{R}^{2}_{q,-}}{1+\mathcal{R}^{2}_{q,-}} \right],\\\nonumber
&=\frac{1}{g\left(\frac{1}{2}-m\right)}\left[1-\frac{g^2 m(1-m)}{1+4g^{2}c^{2}m(1-m)}\right. \\&\left.+\frac{3g^4 m^2(1-m)^2(1-4g^{2}c^{2}m(1-m))}{4(1+4g^{2}c^{2}m(1-m))^2}+\mathcal{O}(g^6)\right]
\end{align}
Once we evaluate the contributions from the turning points, we observed that the terms at the orders $g^2$ and $g^4$ are cancelled. We conjecture again that these quantum corrections from the poles and the turnings points cancel at each order:
 \begin{align}
 \mathcal{I}_{\widehat{\mathcal{V}}}=\frac{1}{g\left( \frac{1}{2}-m\right)}+\mathcal{O}(g^6).
 \end{align}
\subsubsection{The P/NP Relation for Quantum Corrected Elliptic Potential}

Assuming that the quantum corrections at the pole and the turning points to the bilinear pairing residue are cancelling to all order, we can write the P/NP relation for our quantum corrected elliptic potential as
\begin{align}
\boxed{\frac{\partial E}{\partial B}=-g\left[\left(\frac{1}{2}-m\right)g\partial_g-m(1-m)\partial_m \right]_{B,c}\mathcal{A}}
\end{align}
The regularization to get the instanton function $A$ from the non-perturbative period $\mathcal{A}$ is slightly more involved than the examples we studied so far. Let $u_{1}<u_{2}$ be the roots of $N(u)$, then we have two cuts $[-1,u_1]$ and $[u_2,1]$. We choose two counterclockwise contours $C_L$ and $C_R$ enclosing the cuts on the positive sheet, $y\sim i\sqrt{1+2m\xi}u^2$ as $u$ goes to $+\infty$ and define two perturbative quantum periods,
\begin{align}
B_{-}=-\frac{1}{\pi}\oint_{C_L}\lambda, \qquad B_{+}=+\frac{1}{\pi}\oint_{C_L}\lambda.
\end{align} 
On this sheet, the residues at the upper and lower imaginary poles (c.f. Eq. (\ref{poles})) are $-\mathcal{R}_{q,+}/2$ and $+\mathcal{R}_{q,-}/2$, respectively. We can choose a large contour that encircles the cuts and the poles and integrate the WKB differential along it. This integral vanishes since the WKB differential does not have a residue at infinity, we get
\begin{align}
B_{-} - B_{+}=-i \left( \mathcal{R}_{q,+} -\mathcal{R}_{q,-} \right)
\end{align}
Without loss of generality, we choose $B=B_{-}$, which corresponds to a selection of the well whose perturbative spectrum we describe. We introduce the mean action $\mathcal{B}$ by
\begin{align}
\mathcal{B}=\frac{1}{2}(B_{+}+B_{-}),
\end{align}
which we regard as a change of variable. Note that the difference between $B_{\pm}$ is energy independent and preserved by the flow. Consequently, the P/NP equation is exactly the same equation written in different coordinates, namely in $\mathcal{B}$. A non-perturbative cycle connecting two wells contains local matching data from both endpoint wells. We produce two $\Gamma$-function factors and a coupling power determined by the two local actions. This is the relevant general structure in the exact WKB treatment of quantum deformed wells \cite{Kamata:2021jrs}. We can regularize the non-perturbative quantum period $\mathcal{A}$ as
\begin{align}
\mathcal{A}=A+\mathcal{B}+\log\frac{g}{8}+\frac{1}{2}G(B_{-})+\frac{1}{2}G(B_{+}).
\end{align}
In terms of the instanton function $A$ and the mean perturbative action $\mathcal{B}$, our P/NP relation takes the following form
\begin{align}
\frac{\partial E}{\partial\mathcal{B}}=-g\left[ \left(\frac{1}{2}-m\right)\mathcal{B}+\left ( \left(\frac{1}{2}-m\right)g\partial_g -m(1-m)\partial_m\right)_{\mathcal{B},c}A \right].
\end{align}
To the best of our knowledge, this quantum corrected elliptic potential was not studied before in the literature, therefore for completeness we would like to provide the first few terms in the expansion of the perturbative energy $E$ 

\begin{align}\nonumber
E(\mathcal{B},g,m,c)&=\mathcal{B}-g\left(\frac{1}{2}-m\right)\Bigg[\frac{1}{2}\mathcal{B}^2- \frac{1}{2}c^2+\frac{1}{8}\Bigg]-g^2\Bigg[\frac{1}{16}\mathcal{B}^3-\left[\left(\frac{1}{16}+\frac{1}{2}m(1-m) \right)c^2\right.\\\nonumber
&\left. -\frac{3}{64}-\frac{1}{8}m(1-m)\right]\mathcal{B}\Bigg]-g^3\left(\frac{1}{2}-m\right)\Bigg[\frac{5}{64}\mathcal{B}^4-\left( \frac{3}{32}c^2-\frac{17}{128}\right)\mathcal{B}^2\\
&+\frac{1}{1024}(4c^2-1)\left[ (128 m(1-m)+4)c^2 -9\right]\Bigg]+\mathcal{O}(g^4),
\end{align}
and the instanton function $A$
\begin{align}\nonumber
A(\mathcal{B}&,g,m,c)=\frac{S_{cl}(m)}{g}+g\Bigg[ \frac{1}{8}(4c^2-1)m(1-m)S_{cl}(m)+\left( \frac{1}{2}-m\right)\left(\frac{3}{4}\mathcal{B}^2+\frac{3}{16}-\frac{1}{4}c^2\right)\Bigg]\\\nonumber
&+g^2\Bigg[ \left( \frac{5}{32}-\frac{7}{24}m(1-m)\right)\mathcal{B}^3+\Bigg [\frac{17}{128}-\frac{5}{96}m(1-m)-\left( \frac{3}{32}+\frac{3}{8}m(1-m)\right)c^2\Bigg]\mathcal{B}\\
&+\mathcal{O}(g^3).
\end{align}
This P/NP relation is verified up to order $\mathcal{O}(g^4)$.


\section{The Flow of the Scaled Instanton Action}\label{instantonactionflow}

For the potentials without a quantum correction, we made a curious observation about the flow of the scaled instanton action along the Whitham direction is equal to the bilinear pairing residue up to a sign:
\begin{align}\label{instantonflow}
\mathcal{I}_{\widehat{\mathcal{V}}} =-\widehat{\mathcal{V}}\mathcal{S}.
\end{align}
If we check whether this identity still holds for the quantum corrected potential, we might naively think that it could be still satisfied for the potential where the correction term $V_1$ is linear in $z$, because 
\begin{align}
\widehat{\mathcal{V}}\mathcal{S}=-g\partial_g\left(\frac{1}{3g}\right)=-\mathcal{I}_{\widehat{\mathcal{V}}},
\end{align}
Note that the right hand side $\mathcal{I}_{\widehat{\mathcal{V}}}$ was already computed in Eq. (\ref{pairingresiduelinear}). If we make the same comparison between the residue and the flow for the quadratic correction, $V_1=cz^2$, we observe that there is a mismatch between the bilinear pairing residue and the flow of the instanton action. If we take into account the instanton action as $S=1/3$ without any modification due to the quantum correction to the potential. The disagreement can be traced to the additional term $2c/3$ in the residue. It turns out that both the apparent match and mismatch should be more carefully examined since the quantum correction to the potential may change the instanton action and we need to take into account this change when we compare the residue with the flow of the instanton action.

We would like to argue that the equation Eq. (\ref{instantonflow}) is in fact hiding an important piece of information about the potential that will be revealed in a careful study of the instanton action and is essential when we study a quantum corrected potential.  We demonstrate some details for the linear deformation and state the result for the quadratic one. 

We evaluate the saddle action relative to the critical value of the full $g$-dependent potential. This is the same prescription used for complex bions in quantum-modified double-well potentials in \cite{Behtash:2015loa}, c.f. Eq. (5.31),  gives $S_{\rm cl}(g,\alpha)=2\int_{r}^{z_T^{\pm}}\sqrt{2[U(z)-U(r)]}\,dz$ after imposing the critical Euclidean energy, where $r$ denotes the minimum of the potential satifying
\begin{align}
r(r-1)(2r-1)+g\alpha=0,
\end{align}
and $z_T^{\pm}$ are the turning points
\begin{align}
z_T^{\pm}=1-r\pm i\sqrt{2\delta}, \qquad \text{with}\,\,\delta=r(r-1).
\end{align}
We would like to emphasize that the classical instanton action $S_{\rm cl}$ has a $g$-dependence due to the quantum correction $V_1$ to the classical part of the potential $V_0$; i.e. the $g$-dependence does not originate from the derivative corrections of the WKB expansion. These are the critical-energy complex-bion actions studied in the complex-saddle literature \cite{Behtash:2015loa}.

Let us define the frequency $\omega$ at the minimum of the potential,
\begin{align}
\omega^2\equiv U''(r)=6r^2-6r+1.
\end{align}
The exact classical action of the full $g$-dependent potential is given by
\begin{align}
S^{\pm}_{\rm cl}(\omega)&=\frac{\omega}{3}+2g\alpha\log\left [ \frac{\omega+1-2r}{\sqrt{2r(r-1)}}\right]\pm i\pi g\alpha,  \\
&=\frac{1}{3}+g\alpha\left [ 1+\log\frac{2}{g\alpha} \right]\pm i \pi g\alpha+\frac{7}{2}g^2\alpha^2+\mathcal{O}(g^3) ,
\end{align}
showing that the classical instanton action is not just $1/3$ unless $\omega=1$. However, it is easy to see that the frequency $\omega$ gets quantum corrections,
\begin{align}
\omega(g,\alpha)=1+3g\alpha-\frac{21}{2}g^2\alpha^2+\mathcal{O}(g^3).
\end{align} 
We can show that the flow of the scaled instanton action $S_{\rm cl}$ now includes the frequency as well,
\begin{align}
\widehat{\mathcal{V}}\mathcal{S}=-\frac{\omega(g,\alpha)}{3g}.
\end{align}

Following this last observation, we are motivated to propose the following relation, 
\begin{align}\label{instantonflowmodified}
\boxed{\widehat{\mathcal{V}}\mathcal{S}=-\omega(g)\,\mathcal{I}^{\rm cl}_{\widehat{\mathcal{V}}},}
\end{align}
which reproduces our initial observation when $\omega(g)=1$. Frequency dependence of P/NP coefficients also appeared in \cite{Misumi:2025ijd}.

Now we provide further evidence for this relation by considering the quadratic quantum deformation to the potential. For simplicity of presentation, let us call $a=1+2cg$, the potential becomes
\begin{align}
U(z;g)=\frac{1}{2}z^2(z^2-2z+a),
\end{align}
so the modified frequency can be exactly found to be $\omega=\sqrt{a}$. The instanton action can be similarly computed
\begin{align}
S_{\rm cl}(a)=\sqrt{a}\left(1-\frac{2a}{3}\right)+ \frac{1-a}{2}\log\frac{1-\sqrt{a}}{1+\sqrt{a}}.
\end{align}
Using this instanton action, we can verify the validity of Eq. (\ref{instantonflowmodified}) for the squared quantum corrected potential. Thus we have an alternative expression for the P/NP relation for genus-one potentials,
\begin{align}
\boxed{\frac{\partial E}{\partial B}=\omega(g)\frac{ \widehat{\mathcal{V}} \mathcal{A} }{ \widehat{\mathcal{V}}\mathcal{S} } }
\end{align}

\section*{Acknowledgments}
We would like to thank Emir O\u{g}uz Kaya, Eray Konya, Emil Hakan Leeb-Lundberg, Osman Teoman Turgut and Umut Varolg\"{u}ne\c{s} for many inspiring discussions. This work is supported by the Scientific and Technological Research Council of Turkey (TÜBİTAK) under the Grant Number 124F428. We thank TÜBİTAK for their support. This article is based upon work from COST Action 21109 CaLISTA and COST Action 22113 THEORY-CHALLENGES, supported by COST (European Cooperation in Science and Technology). 


\appendix
\section{Appendix: Monodromy in the Deformation Space $\mathcal{B}$}\label{appendixA}

\par{In \cite{Cavusoglu:2024usn}, the generalized P/NP relation is originally motivated by the requirement that the relation must be invariant under the monodromy on the deformation space $\mathcal{B}$. When the deformation parameter $\gamma$ is rotated around $1$ in the case of the deformed quartic potential, the instanton action $S(\gamma)$ changes as }
\begin{align}
S(\gamma)\rightarrow S(\gamma)+2\pi R(\gamma),
\end{align}
which can be immediately seen by the presence of the logarithm in the instanton action. Note that $\gamma=1$ is also the point in the deformation space where the potential is restored to the previously studied (quartic) double well. 

In this Appendix, we want to give a slightly more mathematical account to this observation, since it is tightly connected to the choice of the Whitham flow. We can regard the family of parameter-dependent hyperelliptic curves $\Sigma_{\mathbf{t}}:\,y^2=Q(z;\mathbf{t})$ as a projection 
\begin{align}
\pi: \mathcal{C}\rightarrow \mathcal{B},
\end{align} 
where we denote by $\mathbf{t}\in\mathcal{B}$ the collection of parameters that the curve depends on, such as $E$, $\gamma$ etc. and serve as the coordinates along the base. Over each point $\mathbf{t}$ we have the fiber $\Sigma_{\mathbf{t}}$. The discriminant locus $\Delta\subset\mathcal{B}$ is the set of parameter for which the fiber $\Sigma_{\mathbf{t}}$ is singular. For the family of curves that we are interested in,
\begin{align}
F(z,y;\mathbf{t})\equiv y^2-Q(z;\mathbf{t})=0,
\end{align}
a point of the curve is singular when $F$ satisfies the following conditions,
\begin{align}
F=0,\qquad \partial_y F=0,\qquad \partial_z F=0.
\end{align}
The middle equation above forces $y=0$, thus we end up with 
\begin{align}
Q(z;\mathbf{t})=0,\qquad \partial_z Q(z;\mathbf{t})=0.
\end{align}
Specifically, for the deformed quartic potential, these equations become
\begin{align}
V'(z_c;\gamma)=0, \qquad\mbox{and}\qquad E=V(z_c;\gamma),
\end{align}
which imply that the energy along the locus is equal to the critical values of the potential. Physically, $z_c$ is an equilibrium point of the classical system, and $E$ is its critical energy. Geometrically, they correspond to turning points merging at $z_c$. In many applications, we employ the WKB approach assuming that the turning points are distinct. This assumption fails along the discriminant locus, when a cycle shrinks and the topology of the WKB curve changes.

The discriminant of for the anharmonic oscillator is given by
\begin{align}
\mbox{Disc}_z\,Q=-32E\left[ 64E^2+(54\gamma^4-72\gamma^2+16)E+1-\gamma^2 \right].
\end{align}
The discriminant locus has two components; 
\begin{align}
\Delta=\{E=0\}\cup\{ 64E^2+(54\gamma^4-72\gamma^2+16)E+1-\gamma^2=0\}
\end{align}
Along the first component when $E=0$, when we get
\begin{align}
Q(z;0,\gamma)=-z^2(z^2-2\gamma z+1),
\end{align}
which has a double zero at $z=0$ regardless of the value of the deformation parameter $\gamma$. The double root at $z=0$ indicates the collapse of the turning points, hence the perturbative cycle at the bottom of the well. The WKB curve degenerates to a nodal curve on this component.   

The leading instanton action relevant to the perturbation theory around the minimum of the potential at $z=0$ is obtained in the limit when the energy $E$ is vanishing; i.e. along the first component of the discriminant locus. We normalize the nodal curve to properly describe this limiting non-perturbative period. By defining,
\begin{align}
w=\frac{ip}{z}
\end{align}
the normalized family of curves becomes
\begin{align}
\widehat{\Sigma}_{\gamma}: \qquad w^2=q(z;\gamma)=z^2-2\gamma z+1.
\end{align} 
This is the normalized zero-energy family relevant for the instanton action. Its moving branch points are
\begin{align}
z_{\pm}(\gamma)=\gamma\pm\sqrt{\gamma^2-1},
\end{align}
and the discriminant of the normalized curve is given by 
\begin{align}
\mbox{Disc}_z\,q=4(\gamma^2-1),
\end{align}
hence $\gamma=\pm 1$ are the discriminant points of the normalized $E=0$ family. Away from the discriminant, all branch points remain distinct. Locally in parameter space, the curves are topologically equivalent, and their homology cycles can be transported continuously. However, the complement 
\begin{align}
\mathcal{B}\backslash \Delta
\end{align}
is not simply connected; a cycle will might be transported to another one along a closed loop around a component of the discriminant locus, $\gamma=\pm 1$. Suppose we transport the normalized curve along a loop around $+1$, then the discriminant changes sign and the branch points are exchanged:
\begin{align}
z_{-} \longleftrightarrow z_{+}.
\end{align} 
The instanton path initially ending at $z_{-}$ is transported into the corresponding path ending at $z_{+}$. On the normalized punctured curve, the difference between these two relative paths is the peripheral cycle around infinity, explaining why the instanton action is shifted by the residue of the WKB differential in terms of monodromy around the discriminant locus. Note that the argument presented here was pictorially demonstrated in Figure 1 of  \cite{Cavusoglu:2024usn}.


\section{Appendix: Non-perturbative Quantum Period and Instanton Function}\label{appendixb}
\par{The nonperturbative quantum period includes the the instanton function that enters the exact quantization condition, and they are known to be relates as}
\begin{align}
\mathcal{A}(B,g,R)=A(B,g,R)+a_1(R)B+B\log (gB)+\sum_{k=2}^{\infty}\frac{B_k(1/2)}{k(k-1)}B^{1-k},
\end{align}
where $B_k(x)$ are the Bernoulli polynomials. Here, $a_1(R)$ is not determined from the dual period itself or Whitham flow, but from the $g^0$ normalization mismatch between $\mathcal{A}$ and $A$. Expand the perturbative energy as
\begin{align}
E(B,g,R)=B+gE_1(B,R)+g^2 E_2(B,R)+\mathellipsis.
\end{align}
If we express the residue $\mathcal{I}_{\widehat{\mathcal{V}}}$ coming from the bilinear identity as $\mathcal{I}_{\widehat{\mathcal{V}}}=C(R)/g$, the P/NP relation for quantum periods takes the following form,
\begin{align}
\widehat{\mathcal{V}}\mathcal{A}=-\frac{C(R)}{g}-C(R)\partial_B E_1(B,R)+\mathcal{O}(g).
\end{align}
Along the Whitham flow, we have
\begin{align}
\widehat{\mathcal{V}}B=0,\qquad \widehat{\mathcal{V}}(\mathcal{R}_q)=0,\qquad \mathcal{V}g=g.
\end{align}
Hence, we get 
\begin{align}
\left[ \widehat{\mathcal{V}}\mathcal{A}\right ]_{g^0}&=-C(R)\partial_B E_1(B,R),\\
f(R)\equiv\left[1+R\frac{da_1}{dR}\right]B&=-C(R)\partial_B E_1(B,R),
\end{align}
where we defined $f(R)$ following. We can determine it from the perturbative data and $C(R)$, which allows us to determine $a_1(R)$,
\begin{align}
a_1(R)=\int^{R}\frac{f(\rho)-1}{\rho}d\rho+\mbox{constant}.
\end{align}
The ambiguity coming from the integration constant is actually insignificant along the flow, because it essentially shifts $\mathcal{A}$ by a multiple of $B$, which is killed by the action of the lifted $D$ by construction. Note that the Whitham derivation predicts a nontrivial constraint that the following expression
\begin{align}
-\frac{C(R)}{B}\partial_B E_1(B,R)
\end{align}
depends only on $R$; not $B$. This is actually a beautiful point; $a_1$ can be read off from the first perturbative correction once the local residue coefficient $C(R)$ is known.
Let us go back to the anharmonic oscillator and show that we really get the P/NP relation of \cite{Cavusoglu:2024usn} from the Whitham flow. The first energy correction $E_1$ can be expressed in terms of the deformation parameter $\gamma$ as (Eq. (A1) of \cite{Cavusoglu:2023bai})
\begin{align}
E_1(B,\gamma)=-\frac{3}{4}(5\gamma^2-1)B^2-\frac{1}{16}(7\gamma^2-3),
\end{align} 
which allows us to compute $f(\gamma)$
\begin{align}
f(\gamma)=\frac{5\gamma^2-1}{3\gamma^2-1}.
\end{align}
For completeness, let us also write $a_1(\gamma)$:
\begin{align}
a_1(\gamma)=\log(\gamma^2-1)+\mbox{constant}.
\end{align}
Now we can use the following relation between the nonperturbative period $\mathcal{A}$ and the instanton function $A$,
\begin{align}
\widehat{\mathcal{V}}\mathcal{A}=\widehat{\mathcal{V}}A+f(R)B
\end{align}
to get the Eq. (17) of \cite{Cavusoglu:2024usn}
\begin{align}
\boxed{\frac{\partial E}{\partial B}=-\frac{3}{2}g\left[(5\gamma^2-1)B+(3\gamma^2-1)g\frac{\partial A}{\partial g}+\gamma(\gamma^2-1)\frac{\partial A}{\partial \gamma} \right] .}
\end{align}


\section{Appendix: Exact WKB Approach}\label{appendixc}
In this section, we would like to review the exact WKB method, in particular if the potential has quantum correction terms. We consider the Schr\"{o}dinger equation in the following normalization, 
\begin{align}
\left[-\frac{g^2}{2}\frac{d^2}{dz^2}+U(z;g,\eta) \right]\psi(z)=\xi\psi(z).
\end{align}
The coordinate $z$ is complexified, and $U$ is assumed to be meromorphic in $z$. The parameter $g$ plays the role of the semiclassical expansion parameter. We will make the following WKB Ansatz for the wave function $\psi(z)$,
\begin{align}
\psi(z)=\mathcal{N} \exp\left[\frac{i}{g}\int_{z_0}^{z}\rho(z')dz' \right],
\end{align}
where $\mathcal{N}$ $z$-independent wave function renormalization, and the integration is locally defined with respect to a reference point $z_0$. Substituting this Ansatz into the Schr\"{o}dinger equation yields the so-called Ricatti equation; exact local rewriting of the Schr\"{o}dinger equation,
\begin{align}
\rho^2-ig\rho'=2(\xi-U),
\end{align}
whose formal semiclassical solutions supply the WKB expansion.

Although we do not use potentials with arbitrary orders of $g$ quantum corrections, we will let the potential be an expansion in $g$ itself
\begin{align}
U(u;g,\eta)=\sum_{n\geq0}g^nV_{n}(z;\eta),
\end{align}
and seek a formal expansion solution for $\rho$ taking
\begin{align}
\rho^{\pm}(z;g,\eta,\xi)=\sum_{n\geq0}g^n\rho_{\pm}(z;\eta,\xi).
\end{align}
At the leading order, the Riccati equation gives,
\begin{align}
\left(\rho_{0}^{\pm} \right)^2=2(\xi-V_0), \qquad \rho_{0}^{\pm} =\pm p_0=\pm\sqrt{2(\xi-V_0)}.
\end{align}
These two choices determine two formal branches. Away from the turning points $p_0=0$, the higher coefficients are obtained recursively,
\begin{align}
\rho_n^{\pm}=\frac{1}{2\rho_0^{\pm}}\left[ i(\rho_{n-1}^{\pm})'-\sum_{j=1}^{n-1}\rho_j^{\pm}\rho_{n-j}^{\pm}-2V_n\right],\qquad n\geq1.
\end{align}
We would like to point out the additional term $2V_n$ in the recursion relation as opposed the Riccati recursion relation for potentials without any quantum corrections. Let us look at the first correction on the branch $\rho_{0}^{+}=p_0$ without loss of generality,
\begin{align}
\rho_{1}^{+}=\frac{i}{2}\frac{p'_0}{p_0}-\frac{V_1}{p_0},
\end{align}
where the first term is again a total derivative, but the additional second term due to the quantum corrections to the potential is not, which spoils that the first correction is a total derivative. In the regular WKB, when we integrate this correction over a cycle in the absence of $V_n$'s we do not get any contribution; for the quantum corrected potentials this is not the case anymore. The even and odd splitting with respect to the degree of the correction does not constitute a relevant division. However, we can still define branch-antisymmetric momentum by 
\begin{align}
P\equiv\frac{\rho^{+}-\rho^{-}}{2},
\end{align}
and by subtracting to two Riccati equations at different branches we get 
\begin{align}
\rho^{+}+\rho^{-}=ig\frac{P'}{P},
\end{align}
which in turn implies
\begin{align}\label{rhopm}
\rho^{\pm}=\pm P+\frac{ig}{2}\frac{P'}{P}.
\end{align}
Integrating this expression in the wavefunction Ansatz produces the formal WKB solutions
\begin{align}
\psi^{\pm}(z)=\frac{\mathcal{N}_{\pm}}{\sqrt{P(z)}}\exp\left[ \pm i \int_{z_{0}}^{z}P(z')dz'\right].
\end{align}
In our discussion, we will use the WKB differential $\lambda$ governing the quantum actions,
\begin{align}
\lambda=\frac{1}{g}Pdz.
\end{align} 
For a $g$-independent potential, $P$ contains only even powers of $g$ in the expansion at fixed $\xi$, and we recover the more familiar WKB expansion. 

Finally, we can substitute Eq. (\ref{rhopm}), which relates $\rho^{\pm}$ to $P$, into the Riccati equation and obtain the formula we frequently make us of,
\begin{align}\label{riccattifull}
P^2+\frac{g^2}{2}\frac{P''}{P}-\frac{3g^2}{4}\left( \frac{P'}{P}\right)^2=2(\xi-U).
\end{align}

The series constructed above are generally asymptotic. Exact WKB supplements the formal recursion with Borel summation, under suitable analyticity and summability assumptions, to obtain actual solutions in appropriate sectors. Their continuation across Stokes curves is described by connection formulas, together with the boundary conditions, these determine the quantization conditions.

\bibliography{refs}

@book{Ecalle:1981,
    author = "\'Ecalle, Jean",
    title = "{Les Fonctions R\'esurgentes}",
    publisher = "Publications Math\'ematiques d'Orsay",
    year = "1981",
    volume = "I-III",
    address = "Orsay, France"
}

@article{vanSpaendonck:2023znn,
    author = "van Spaendonck, Alexander and Vonk, Marcel",
    title = "{Exact instanton transseries for quantum mechanics}",
    eprint = "2309.05700",
    archivePrefix = "arXiv",
    primaryClass = "hep-th",
    doi = "10.21468/SciPostPhys.16.4.103",
    journal = "SciPost Phys.",
    volume = "16",
    number = "4",
    pages = "103",
    year = "2024"
}

@article{Cavusoglu:2024usn,
    author = "{\c{C}}avu{\c{s}}o{\u{g}}lu, Atakan and Koz{\c{c}}az, Can and Tezgin, Kemal",
    title = "{Resurgence of deformed genus-1 curves: A novel perturbative/nonperturbative relation}",
    eprint = "2408.02628",
    archivePrefix = "arXiv",
    primaryClass = "hep-th",
    doi = "10.1103/PhysRevD.111.105007",
    journal = "Phys. Rev. D",
    volume = "111",
    number = "10",
    pages = "105007",
    year = "2025"
}

@article{Ture:2026ktb,
    author = {T{\"u}re, Mustafa and {\"U}nsal, Mithat},
    title = "{Quantum geometry at arbitrary genus - I: Anharmonic potentials}",
    eprint = "2610.10940",
    archivePrefix = "arXiv",
    primaryClass = "hep-th",
    month = "10",
    year = "2026"
}

@article{upcoming,
    author = "Kaya, Emir O\u{g}uz and Konya, Eray and Koz\c{c}az, Can",
    title = "{work in progress}",
    eprint = "",
    archivePrefix = "",
    primaryClass = "",
    doi = "",
    journal = "",
    volume = "",
    number = "",
    pages = "",
    year = "",
}

@article{Fischbach:2018yiu,
    author = "Fischbach, Fabian and Klemm, Albrecht and Nega, Christoph",
    title = "{WKB method and quantum periods beyond genus one}",
    eprint = "1803.11222",
    archivePrefix = "arXiv",
    primaryClass = "hep-th",
    doi = "10.1088/1751-8121/aae8b0",
    journal = "J. Phys. A",
    volume = "52",
    number = "7",
    pages = "075402",
    year = "2019"
}

@article{DDP,
    author = {Delabaere, Eric and Dillinger, Hervé and Pham, Frédéric},
    title = "{Exact semiclassical expansions for one-dimensional quantum oscillators}",
    journal = {J. Math. Phys. (N.Y.)},
    volume = {38},
    number = {12},
    pages = {6126-6184},
    year = {1997},
    month = {12},
    issn = {0022-2488},
    doi = {10.1063/1.532206},
    url = {https://doi.org/10.1063/1.532206},
}

@article{Voros1983,
author = {Voros, A.},
journal = {Annales de l'I.H.P. Physique théorique},
number = {3},
pages = {211-338},
publisher = {Gauthier-Villars},
title = "{The return of the quartic oscillator. The complex WKB method}",
url = {http://eudml.org/doc/76217},
volume = {39},
year = {1983},
}

@article{Alvarez2004,
    author = {\'{A}lvarez, Gabriel},
    title = "{Langer–Cherry derivation of the multi-instanton expansion for the symmetric double well}",
    journal = {J. Math. Phys. (N.Y.)},
    volume = {45},
    number = {8},
    pages = {3095-3108},
    year = {2004},
    month = {07},
    issn = {0022-2488},
    doi = {10.1063/1.1767988},
    url = {https://doi.org/10.1063/1.1767988},
}

@article{Dunne:2025mye,
    author = "Dunne, Gerald V.",
    title = "{Introductory lectures on resurgence}",
    eprint = "2511.15528",
    archivePrefix = "arXiv",
    primaryClass = "hep-th",
    doi = "10.1140/epjs/s11734-026-02135-y",
    journal = "Eur. Phys. J. ST",
    volume = "235",
    number = "17",
    pages = "3891--3938",
    year = "2026"
}

@article{Aniceto:2018bis,
    author = "Aniceto, In\^es and Ba\c{s}ar, G{\"o}k\c{c}e and Schiappa, Ricardo",
    title = "{A primer on resurgent transseries and their asymptotics}",
    eprint = "1802.10441",
    archivePrefix = "arXiv",
    primaryClass = "hep-th",
    reportNumber = "NSF-ITP-17-153",
    doi = "10.1016/j.physrep.2019.02.003",
    journal = "Phys. Rep.",
    volume = "809",
    pages = "1--135",
    year = "2019"
}

@article{MarinoBook1,
    author = "Mari\~no, Marcos",
    title = "{Advanced topics in quantum mechanics }",
    journal = "Cambridge University Press",
    year = "2021"
}

@article{MarinoBook2,
    author = "Mari\~no, Marcos",
    title = "{Instantons and large N}",
    journal = "Cambridge University Press",
    year = "2015"
}

@article{Alvarez:2000,
doi = {10.1088/0305-4470/33/29/302},
url = {https://dx.doi.org/10.1088/0305-4470/33/29/302},
year = {2000},
volume = {33},
number = {29},
pages = {5171},
author = {Gabriel Álvarez and Carmen Casares},
title = {Exponentially small corrections in the asymptotic expansion of the eigenvalues of the cubic anharmonic oscillator},
journal = {J. Phys. A}
}

@article{Zinn-Justin:2004vcw,
    author = "Zinn-Justin, Jean and Jentschura, Ulrich D.",
    title = "{Multi-instantons and exact results I: Conjectures, WKB expansions, and instanton interactions}",
    eprint = "quant-ph/0501136",
    archivePrefix = "arXiv",
    doi = "10.1016/j.aop.2004.04.004",
    journal = "Ann. Phys. (N.Y.)",
    volume = "313",
    pages = "197--267",
    year = "2004"
}

@article{Zinn-Justin:2004qzw,
    author = "Zinn-Justin, Jean and Jentschura, Ulrich D.",
    title = "{Multi-instantons and exact results II: Specific cases, higher-order effects, and numerical calculations}",
    eprint = "quant-ph/0501137",
    archivePrefix = "arXiv",
    doi = "10.1016/j.aop.2004.04.003",
    journal = "Ann. Phys. (N.Y.)",
    volume = "313",
    pages = "269--325",
    year = "2004"
}

@article{Jentschura:2010zza,
    author = "Jentschura, Ulrich D. and Surzhykov, Andrey and Zinn-Justin, Jean",
    title = "{Multi-instantons and exact results III: Unification of even and odd anharmonic oscillators}",
    eprint = "1001.3910",
    archivePrefix = "arXiv",
    primaryClass = "math-ph",
    doi = "10.1016/j.aop.2010.01.002",
    journal = "Ann. Phys. (N.Y.)",
    volume = "325",
    pages = "1135--1172",
    year = "2010"
}

@article{Dunne:2013ada,
    author = {Dunne, Gerald V. and \"Unsal, Mithat},
    title = "{Generating nonperturbative physics from perturbation theory}",
    eprint = "1306.4405",
    archivePrefix = "arXiv",
    primaryClass = "hep-th",
    doi = "10.1103/PhysRevD.89.041701",
    journal = "Phys. Rev. D",
    volume = "89",
    number = "4",
    pages = "041701",
    year = "2014"
}

@article{Dunne:2014bca,
    author = {Dunne, Gerald V. and {\"U}nsal, Mithat},
    title = "{Uniform WKB, multi-instantons, and resurgent trans-series}",
    eprint = "1401.5202",
    archivePrefix = "arXiv",
    primaryClass = "hep-th",
    doi = "10.1103/PhysRevD.89.105009",
    journal = "Phys. Rev. D",
    volume = "89",
    number = "10",
    pages = "105009",
    year = "2014"
}

@InProceedings{Dunne:2016qix,
    author="Dunne, Gerald V.
and {\"U}nsal, Mithat",
    editor="Fauvet, Fr{\'e}d{\'e}ric
and Manchon, Dominique
and Marmi, Stefano
and Sauzin, David",
    title="{WKB and resurgence in the Mathieu equation}",
    booktitle="Resurgence, Physics and Numbers",
    year="2017",
    publisher="Scuola Normale Superiore",
    address="Pisa",
    pages="249--298",
    isbn="978-88-7642-613-1",
    eprint = "1603.04924",
    archivePrefix = "arXiv",
    primaryClass = "math-ph",
    doi = "10.1007/978-88-7642-613-1_6"
}

@article{Dunne:2016jsr,
    author = "Dunne, Gerald V. and {\"U}nsal, Mithat",
    title = "{Deconstructing zero: Resurgence, supersymmetry and complex saddles}",
    eprint = "1609.05770",
    archivePrefix = "arXiv",
    primaryClass = "hep-th",
    doi = "10.1007/JHEP12(2016)002",
    journal = "J. High Energy Phys.",
    volume = "2016",
    number = "12",
    pages = "2",
    year = "2016"
}

@article{Misumi:2025ijd,
    author = "Misumi, Tatsuhiro and Pazarba{\c{s}}{\i}, Cihan",
    title = "{Exact WKB in all sectors. Part II. Potentials with non-degenerate saddles}",
    eprint = "2511.20778",
    archivePrefix = "arXiv",
    primaryClass = "hep-th",
    doi = "10.1007/JHEP05(2026)306",
    journal = "JHEP",
    volume = "05",
    pages = "306",
    year = "2026"
}

@article{Kamata:2021jrs,
    author = {Kamata, Syo and Misumi, Tatsuhiro and Sueishi, Naohisa and {\"U}nsal, Mithat},
    title = "{Exact WKB analysis for SUSY and quantum deformed potentials: Quantum mechanics with Grassmann fields and Wess-Zumino terms}",
    eprint = "2111.05922",
    archivePrefix = "arXiv",
    primaryClass = "hep-th",
    doi = "10.1103/PhysRevD.107.045019",
    journal = "Phys. Rev. D",
    volume = "107",
    number = "4",
    pages = "045019",
    year = "2023"
}

@article{Behtash:2015loa,
    author = {Behtash, Alireza and Dunne, Gerald V. and Sch\"afer, Thomas and Sulejmanpasic, Tin and \"Unsal, Mithat},
    title = "{Toward Picard\textendash{}Lefschetz theory of path integrals, complex saddles and resurgence}",
    eprint = "1510.03435",
    archivePrefix = "arXiv",
    primaryClass = "hep-th",
    doi = "10.4310/AMSA.2017.v2.n1.a3",
    journal = "Ann. Math. Sci. Appl.",
    volume = "02",
    pages = "95--212",
    year = "2017"
}

@article{Dorigoni:2014hea,
    author = "Dorigoni, Daniele",
    title = "{An Introduction to Resurgence, Trans-Series and Alien Calculus}",
    eprint = "1411.3585",
    archivePrefix = "arXiv",
    primaryClass = "hep-th",
    reportNumber = "DAMTP-2014-44",
    doi = "10.1016/j.aop.2019.167914",
    journal = "Annals Phys.",
    volume = "409",
    pages = "167914",
    year = "2019"
}

@article{Basar:2017hpr,
    author = "Ba\c{s}ar, G{\"o}k\c{c}e and Dunne, Gerald V. and {\"U}nsal, Mithat",
    title = "{Quantum geometry of resurgent perturbative/nonperturbative relations}",
    eprint = "1701.06572",
    archivePrefix = "arXiv",
    primaryClass = "hep-th",
    doi = "10.1007/JHEP05(2017)087",
    journal = "J. High Energy Phys.",
    number = "5",
    pages = "87",
    year = "2017",
    volume = "2017"
}

@article{Kozcaz:2016wvy,
    author = {Koz\c{c}az, Can and Sulejmanpasic, Tin and Tanizaki, Yuya and \"Unsal, Mithat},
    title = "{Cheshire Cat resurgence, self-resurgence and quasi-exact solvable systems}",
    eprint = "1609.06198",
    archivePrefix = "arXiv",
    primaryClass = "hep-th",
    reportNumber = "RBRC-1294",
    doi = "10.1007/s00220-018-3281-y",
    journal = "Commun. Math. Phys.",
    volume = "364",
    number = "3",
    pages = "835--878",
    year = "2018"
}

@article{Bender:1969si,
    author = "Bender, Carl M. and Wu, Tai Tsun",
    title = "{Anharmonic oscillator}",
    doi = "10.1103/PhysRev.184.1231",
    journal = "Phys. Rev.",
    volume = "184",
    pages = "1231--1260",
    year = "1969"
}

@article{Bender:1973rz,
    author = "Bender, Carl M. and Wu, T. T.",
    title = "{Anharmonic oscillator. II: A study of perturbation theory in large order}",
    doi = "10.1103/PhysRevD.7.1620",
    journal = "Phys. Rev. D",
    volume = "7",
    pages = "1620--1636",
    year = "1973"
}

@article{Cavusoglu:2023bai,
    author = "\c{C}avu\c{s}o\u{g}lu, Atakan and Koz\c{c}az, Can and Tezgin, Kemal",
    title = "{Unified genus-1 potential and a parametric perturbative/nonperturbative relation}",
    eprint = "2311.17850",
    archivePrefix = "arXiv",
    primaryClass = "hep-th",
    doi = "10.1103/PhysRevD.110.125001",
    journal = "Phys. Rev. D",
    volume = "110",
    number = "12",
    pages = "125001",
    year = "2024"
}

@article{Gorsky:2014lia,
    author = "Gorsky, A. and Milekhin, A.",
    title = "{RG-Whitham dynamics and complex Hamiltonian systems}",
    eprint = "1408.0425",
    archivePrefix = "arXiv",
    primaryClass = "hep-th",
    reportNumber = "ITEP-TH-19-14",
    doi = "10.1016/j.nuclphysb.2015.03.028",
    journal = "Nucl. Phys. B",
    volume = "895",
    pages = "33--63",
    year = "2015"
}
\end{document}